\documentclass[aps,prb,
preprint,
showpacs,
floatfix,
]{revtex4-1}
\usepackage{bm}
\usepackage{amsmath}    
\usepackage{graphicx}   
\usepackage[hang]{subfigure}
\newcommand{\jun}{junction }
\newcommand{\juns}{junctions }
\newcommand{\Jos}{Josephson }
\begin{document}
\title {Long Josephson Tunnel Junctions with Doubly Connected Electrodes}
\thanks{Published in Phys. Rev. B. (April 2012)}
\author{R. Monaco}
\affiliation{Istituto di Cibernetica del CNR, Comprensorio Olivetti, 80078 Pozzuoli, Italy and Facolt$\grave{a}$ di Scienze, Universit$\grave{a}$ di Salerno, 84084 Fisciano, Italy}\email
{roberto.monaco@cnr.it}
\author{J. Mygind}
\affiliation{DTU Physics, B309, Technical University of Denmark, DK-2800 Lyngby, Denmark}
\author{V.\ P.\ Koshelets}
\affiliation{Kotel'nikov Institute of Radio Engineering and Electronics,
Russian Academy of Science, Mokhovaya 11, Bldg 7, 125009 Moscow, Russia.}
\date{\today}
\begin{abstract}
In order to mimic the phase changes in  the primordial Big Bang, several {\it cosmological} solid-state experiments have been conceived, during the last decade, to investigate the spontaneous symmetry breaking in superconductors and superfluids cooled through their transition temperature. In one of such experiments the number of magnetic flux quanta spontaneously trapped in a superconducting loop was measured by means of a long Josephson tunnel junction built on top of the loop itself. We have analyzed this system and found a number of interesting features not occurring in the conventional case with simply connected electrodes. In particular, the fluxoid quantization results in a frustration of the Josephson phase, which, in turn, reduces the junction critical current. Further, the possible stable states of the system are obtained  by a self-consistent application of the principle of minimum energy.  The theoretical findings are supported by measurements on a number of  samples having different geometrical configuration. The experiments demonstrate that a very large signal-to-noise ratio can be achieved in the flux quanta detection.
\end{abstract}
\maketitle
\tableofcontents
\listoffigures

\section{INTRODUCTION}

\begin{figure}[tb]
\centering
\subfigure[ ]{\includegraphics[width=7.0cm]{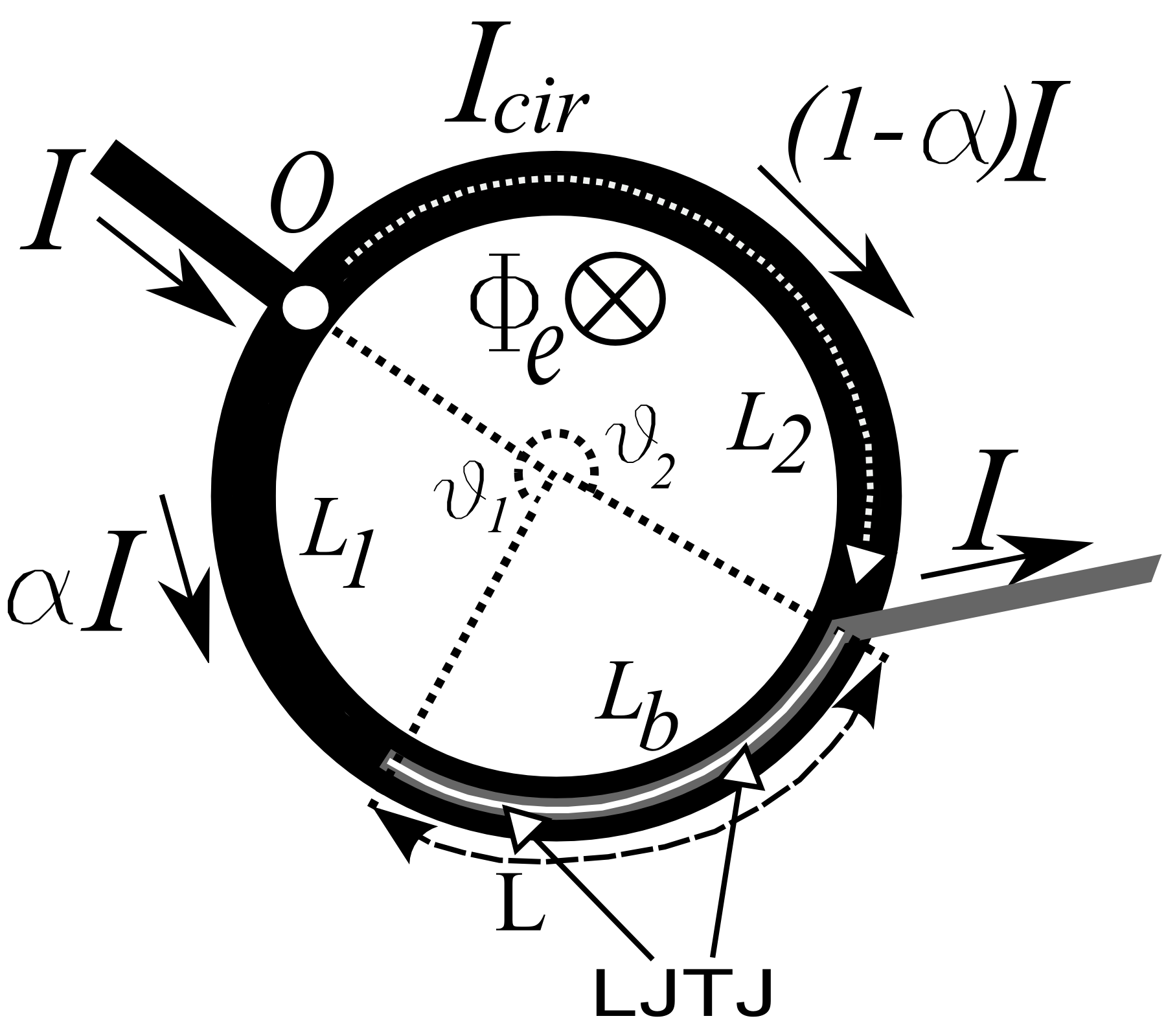}}
\subfigure[ ]{\includegraphics[width=7.0cm]{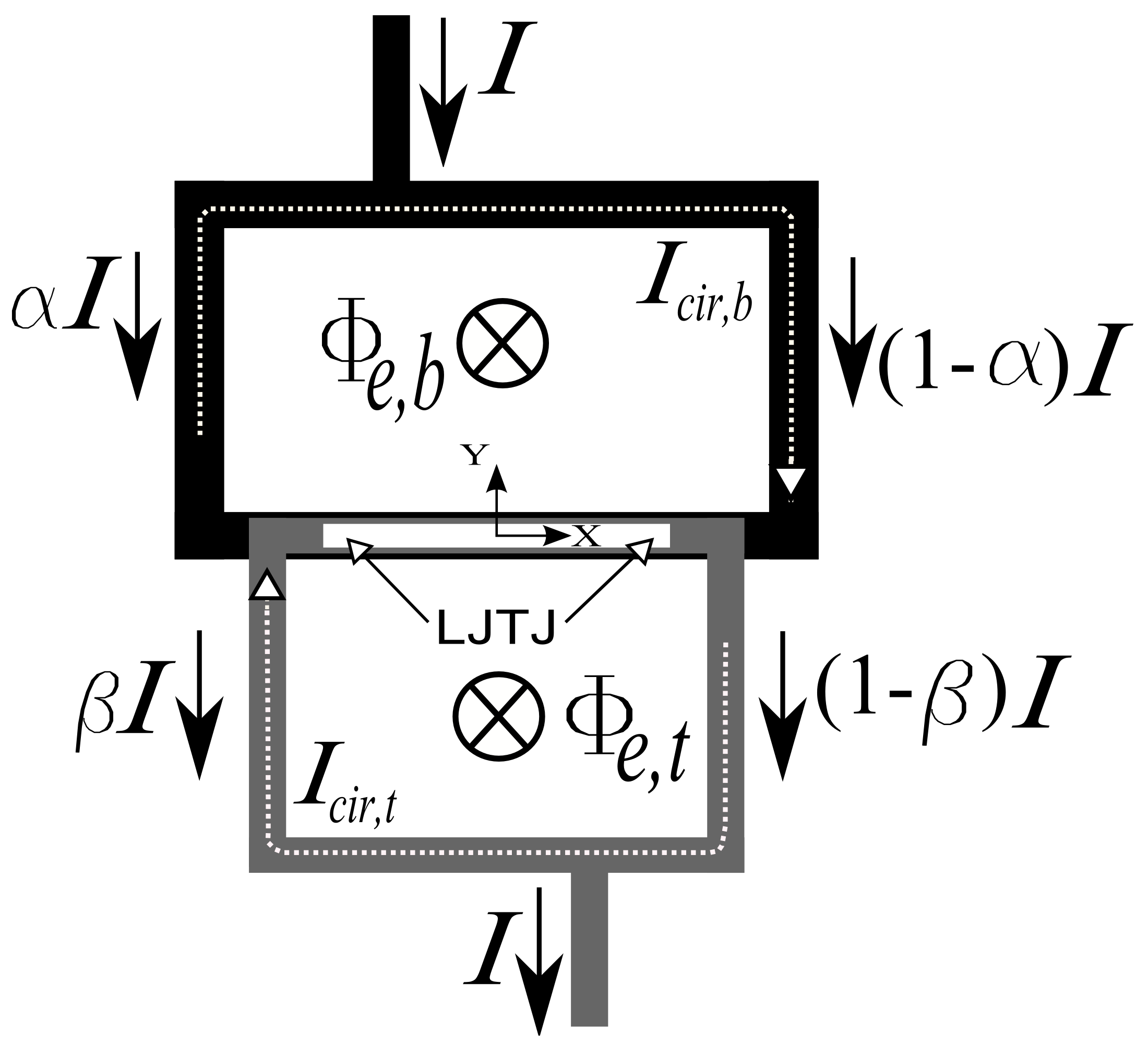}}

\caption{ a) Sketch of an in-line Josephson tunnel junction with a single doubly connected ring-shaped base electrode. $L_{loop}=L_1+L_2+L_b$. b) Sketch of an in-line Josephson tunnel junction with two doubly connected electrodes. The base electrode is in black, the top electrode is in gray, and the tunneling insulating layer is white. The dotted withe arrows indicate the direction of the circulating currents.\label{geometry}}

\end{figure}

Long Josephson tunnel junctions (LJTJs) were traditionally used to investigate the physics of non-linear phenomena\cite{barone}. In the last decade they have been employed to shed light on other fundamental concepts in physics such as the symmetry principles and how they are broken\cite{PRLS,PRB08,gordeeva}. A recent experiment\cite{PRB09} has demonstrated  spontaneous symmetry breaking during the superconducting phase transition of a metal ring and both fluxoids or antifluxoids can be trapped in the ring while it is cooled rapidly through the superconducting critical temperature. The basic phenomenon of quantization of magnetic flux in a multiply connected superconductor was suggested long time ago as one among several possible condensed matter \textit{cosmological} experiments\cite{zurek2} suitable to check the validity of the \textit{causality} principle in the early Universe\cite{kibble1}. In the experiment of Ref.\cite{PRB09} the magnetic flux quanta are spontaneously trapped in the ring during its cooling through the transition temperature.  Much later at lower temperature when superconductivity is fully established, the number $0,\pm 1, \pm 2$ ... of flux quanta is registered as a function of the quench rate.  This can be done in a variety of ways, one of which is the detection of the induced persistent currents by the magnetic field modulation of the critical current of a planar LJTJ \textit{built} on top of the ring. In these experiments, the quench rate can be varied over four decades. This allows for an accurate check of the theoretical predictions of the involved second-order phase transitions. This is of  interest within cosmology and of major importance for the physical understanding of many order-disorder processes. However, the working principles of that experiment had not yet been reported. The general task of this work is to study the static properties of a planar LJTJ for which at least one of the superconducting electrodes is multiply-connected, i.e., not every closed path can be transformed into a point. In the simplest case, one of the superconducting thin-film  stripes forming the LJTJ is shaped as a doubly-connected loop. This configuration is illustrated in Fig.\ref{geometry}(a) in which the ring-shaped base electrode is in black, while the top electrode is in gray and the junction area is in white. The geometry of the loop is not critical to our discussion; however, a ring-shaped bottom electrode simplifies the analysis. 

\noindent For the sake of generality, in our analysis, we will include an external flux $\Phi_e$ linked to the loop by some externally applied field and the presence of an integer number $n$ of flux quanta trapped in the loop; altogether they induce a current $I_{cir}=(n\Phi_0-\Phi_e)/L_{loop}$ circulating clockwise in the loop and inversely proportional to its inductance, $L_{loop}$; in turns, the circulating current produces at the loop surface a radial magnetic field, $H_{rad} \propto I_{cir}$, that adds to any external  field, $H_{app}$, applied in the loop plane. With no loss of generality, we will assume that the width, $W_t$, of the top film does not exceed the width, $W_b$, of the bottom one, $W_t\leq W_b$ and, to simplify the analysis, we will also assume that both widths are much smaller than the mean radius, $R$, of the ring; in this narrow ring approximation, the current distribution in the ring and the surface radial field become radially independent\cite{brandt}. A dc current $I$ is injected into the loop at an arbitrary point $O$ along the ring and is inductively split in the two loop arms before going through the LJTJ; let $\alpha$ (1-$\alpha$) be the fraction of the bias current $I$ diverted in the left (right) side of the loop. In principle, $\alpha$ values outside the $[0,1]$ range are possible if the current $I$ would include also a persistent current $I_{cir}$ circulating in the loop: however, since the two currents are independent, we will treat them separately. Independently of the $\alpha$ value, the bias current $I$ is extracted at one end of the junction via the top electrode. With the current entering and exiting at the junction extremities we have the well-known case of the so-called \textit{in-line} configuration treated in the pioneering works on LJTJs soon after the discovery of the Josephson effect\cite{ferrel,OS,stuehm,basa,radparvar85}.

\noindent Throughout the paper we will limit our interest to LJTJs in the zero-voltage time-independent state; this can be achieved as far as the applied current $I$ is smaller than the junction critical current $I_c$. To further simplify the analysis, we assume that the Josephson current density $J_c$ is uniform over the barrier area and that the junction width $W_{}$ is smaller than the Josephson penetration depth $\lambda _J\equiv \sqrt{\Phi_0/ 2\pi \mu _{0}d_{e} J_{c}}$ setting the length unit of the physical processes occurring in the \Jos \jun  (here $\Phi_0$ is the magnetic flux quantum, $\mu_0$ the vacuum permeability and $d_e$ the junction magnetic thickness). The gauge-invariant phase difference $\phi$ of the order parameters of the superconductors on each side of the tunnel barrier obeys the Josephson equations\cite{joseph}:

\begin{equation}
\label{jos}
J_Z(X)=J_c \sin \phi(X),
\end{equation}

\noindent and

\begin{equation}
\label{gra}
\kappa {\bf \nabla} \phi(X) = {\bf H}\times {\bf \hat{n}},
\end{equation}

\noindent in which $-{\rm{L}}/2\leq X \leq {\rm{L}}/2$ is a curvilinear coordinate and ${\rm{L}}$ is the long dimension of the junction. The net current crossing the tunnel barrier is $I \equiv  W_{} \int_{-{\rm{L}}/2}^{{\rm{L}}/2} J_Z(X)dX$. The last equation states that the phase gradient is everywhere proportional to the local magnetic field ${\bf H}$ and parallel to the barrier plane. Therefore, in the case of a curvilinear one-dimensional junction, a uniform external field applied in the junction plane has to be replaced by its radial component\cite{PRB96}. $\kappa \equiv {\Phi_0}/{2\pi d_e \mu _0}=J_c \lambda_J^2$ has the dimension of a current ($\kappa\approx 2.5\,$mA when $d_e \approx 100\,$nm, which is typical of all-Niobium \Jos junctions) and ${\bf \hat{n}}$ is the versor normal to the insulating barrier separating the two superconducting electrodes. It is well known\cite{ferrel,OS} that combining Eqs.(\ref{jos}) and (\ref{gra}) with the static Maxwell's equations, a static sine-Gordon equation is obtained that describe the behavior of a one-dimensional in-line LJTJ:

\begin{equation}
\lambda_J^2 \frac{d^2 \phi}{d X^2} = \sin \phi(X).
\label{sG}
\end{equation}

\noindent  Equation(\ref{sG}) was first introduced in the analysis of \textit{asymmetric} in-line LJTJs by Ferrel and Prange\cite{ferrel} in 1963; few years later, Owen and Scalapino\cite{OS} reported an extensive study of its analytical solutions for \textit{symmetric} in-line \Jos junctions (provided that ${\rm{L}}\geq\pi \lambda_J /2$). Ampere's law applied along the barrier perimeter requires that the magnetic fields at the two ends of the junctions differ by the amount of the enclosed current: $I=W_{} [H_Y ({\rm{L}}/2) -H_Y (-{\rm{L}}/2)]$. We remark that Eqs.(\ref{jos}), (\ref{gra}) and (\ref{sG}) automatically satisfy the Ampere's law. 

\noindent As it is usually done in the modeling of Josephson interferometers, it is useful to divide the loop inductance $L_{loop}$ in three inductive paths characterized by positive coefficients $L_1$, $L_2$ and $L_b$ having units of inductance such that $L_{loop}=L_1+L_2+L_b$ so that any current $I_{cir}$ circulating around the loop \textit{sees} them in series; furthermore, $\alpha=1$ ($\alpha=0$) in the limit of $L_2>>L_1+L_b$ ($L_2<<L_1+L_b$). More specifically, $L_{1,2}$ is the angular fraction $\vartheta_{1,2}/2\pi$ of the inductance for an isolated superconducting narrow ($W_b<<R$) ring\cite{brandt} $L_{ring}=\mu_0 R (\ln 8R/W_b - 2 + \ln 4)$ and $L_b$, that should not be mistaken as the inductance of the LJTJ, is given by the junction physical length  ${\rm{L}}$ times the inductance per unit length of the junction bottom strip-line $\mathcal{L}_b$, related to the magnetic energy stored within a London penetration distance of its surface. If the bottom and top superconducting films have, respectively, thickness $d_{b}$ and $d_{t}$ and bulk London penetration depths $\lambda_{Lb}$ and $\lambda_{Lt}$, then, in presence of a quasi-static magnetic field, their effective penetration depth\cite{wei} is $\lambda_{b,t} =\lambda_{Lb,t} \tanh (d_{b,t}/2 \lambda_{Lb,t})$ which reduces to $\lambda_{Lb,t}$ in the case of thick superconducting films ($d_{b,t} >5 \lambda_{Lb,t}$).  The magnetic penetration $d_e$ of a tunnel barrier with negligible height $t_{ox}<< d_{b,t}$ is\cite{wei} $d_e \simeq \lambda_{b} + \lambda_{t}$. Insofar as the width $W_b$ is much larger\cite{swihart,orlando} than the strip-line magnetic thickness $d_e$, then $\mathcal{L}_b \simeq \mu_0 \lambda_{b}/W_b$; this expression also takes into account the kinetic inductance due to the motion of the superelectrons. In the wide-strip approximation, most of the magnetic energy is confined in the region between the plates and the fringing field can be ignored; as the strip width becomes narrower, the fringe field effects become more important and may dominate if $W_b$ and $d_e$ are comparable\cite{chang}. It is worth pointing out that, since, in all practical cases, the width of the loop is much larger than the London penetration depth, then $\mathcal{L}_b$ is considerably smaller than the inductance per unit length along the ring $L_{ring}/2\pi R$; this is due to the presence of a counter electrode acting as a superconducting ground plane\cite{chang,vanDuzer}. Similarly, we introduce $\mathcal{L}_t \simeq \mu_0 \lambda_{t}/W_t$, the inductance per unit length along the current direction of the top plate. Since the electrodes have different widths and penetration depths, in general, $\mathcal{L}_t \neq \mathcal{L}_b$. In Section III we will show that for our high quality all-Niobium LJTJs, having the base electrode thinner and wider than the top one, we found $\mathcal{L}_t \approx 3 \mathcal{L}_b$. According to the theory of the two-conductor transmission lines\cite{kraus}, the inductance per unit length, $\mathcal{L}_J$, of a LJTJ, seen as a transmission line structure, is simply obtained as the sum of the inductances/unit lengths of the bottom and top stripes, i.e., 

\begin{equation}
\mathcal{L}_J=\mathcal{L}_b + \mathcal{L}_t=\mu_0 \left( \frac{\lambda_{b}}{W_b}+ \frac{\lambda_{t}}{W_t} \right).
\label{indu}
\end{equation}

\noindent Historically, the boundary conditions for Eq.(\ref{sG}) were derived under the implicit assumption that $W_b = W_t=W_{}$, so that\cite{scott76,vanDuzer} $\mathcal{L}_J=\mu_0 d_e/W_{}$. However these conditions are not fulfilled in real samples, especially for window-type LJTJs used nowadays whose electrodes have quite different widths; typically, $W_b>W_t>W_{}$. 

\noindent The paper is organized in the following way. In Section II we will overcome this limitation by extending the existing theoretical model\cite{OS} to LJTJs having different electrodes widths, strictly speaking, different inductances per unit length. At the same time we will derive the most general boundary conditions for Eq.(\ref{sG}) needed to correctly describe any self-field effect in LJTJs. Next we will focus on the specific case of LJTJs with doubly connected electrode(s). Later on we will consider the consequences of the fluxoid quantization and energy minimization principles. In the next section we will describe our experimental setup and our samples; in addition we will present their magnetic diffraction patterns and discuss how the experimental findings can be unambiguously interpreted in term of our  modeling.  Finally, the conclusions will be drawn in Section IV.

\section{THEORY}

\begin{figure}[tb]
\centering
\subfigure[ ]{\includegraphics[width=7.0cm]{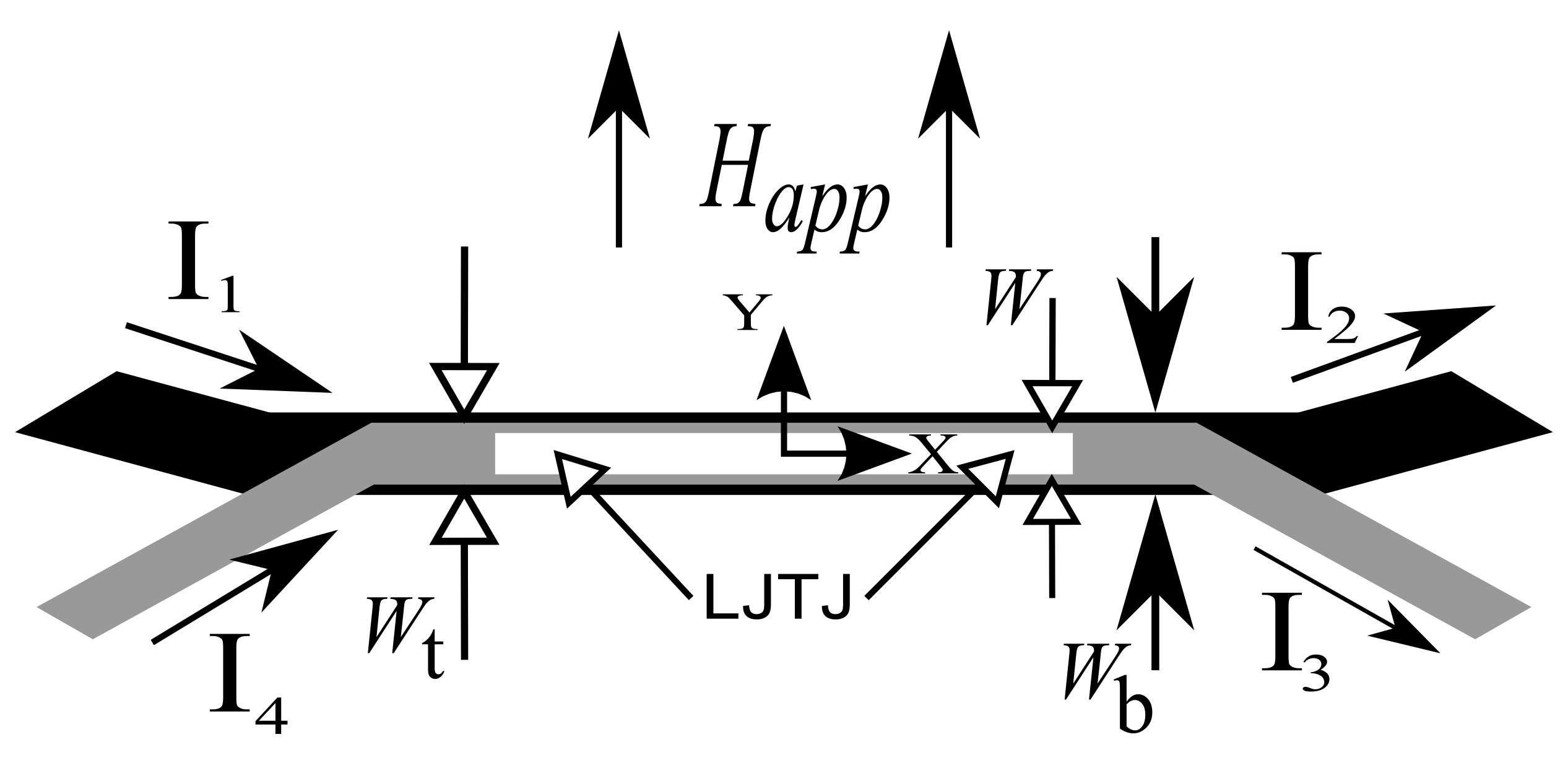}}
\subfigure[ ]{\includegraphics[width=7.0cm]{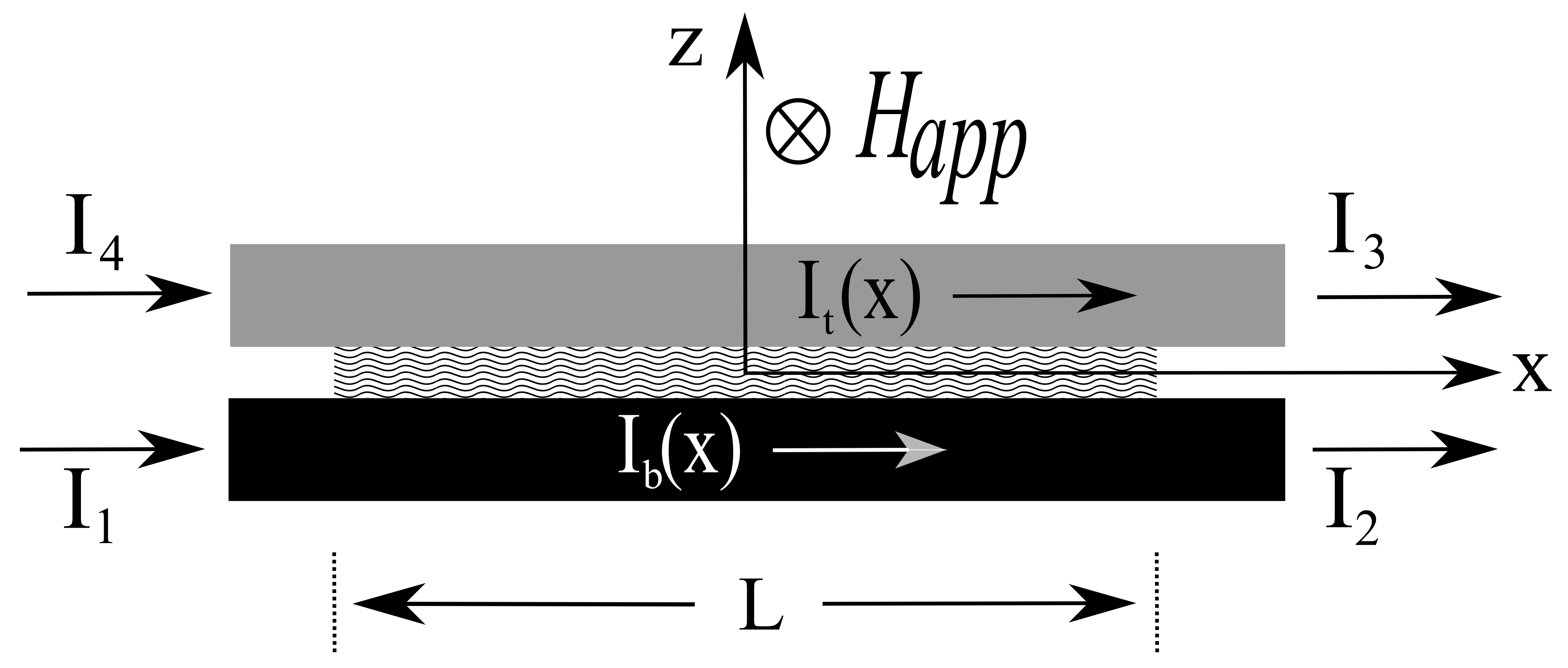}}
\caption{(a) Top view (not to scale) of an in-line long Josephson tunnel junction in the most general bias configuration. (b) Cross section (not to scale) of the junction area showing the currents in the electrodes. The base electrode is in black, the top electrode in gray and the tunneling insulating layer is white in (a) while it has a wavy pattern in (b).\label{view}}
\end{figure}

\noindent Figure\ref{view}(a) displays the top view of an in-line LJTJ having the most general biasing configuration. Only three of the four currents $I_i$ indicated in the figure are independent, since the charge conservation requires that $I_1+I_4=I_2+I_3$. In addition, the $I_i$'s can be expressed in terms of the net current $I$ crossing the tunnel barrier as $I=I_1-I_2=I_3-I_4$. The junction cross section is sketched in Fig.\ref{view}(b) together with the current distribution at the input and output end of the \Jos structure and along the \jun electrodes. Here, $I_b(X)$ and $I_t(X)$ denote, respectively, the local supercurrent flowing parallel to the insulating layer in the bottom and top junction electrodes with $X \in [-{\rm{L}}/2, {\rm{L}}/2]$, so that $I_1=I_b(-{\rm{L}}/2)$, $I_2=I_b({\rm{L}}/2)$, $I_3=I_t({\rm{L}}/2)$, and $I_4=I_t(-{\rm{L}}/2)$. Next we observe that, due to the charge conservation, the total current, $I_b(X) +I_t(X)$, through any junction cross section in the $Y-Z$ plane must be constant.  For in-line LJTJs, it is important to distinguish between the symmetric and fully asymmetric configurations: in the former, the bias current $I$ enters at one extremity and exits at the other\cite{OS,stuehm,radparvar81} ($I=\pm I_1=\pm I_3$ and $I_2=I_4=0$), while in the latter, the bias current enters and exits from the same extremity\cite{ferrel,stuehm,basa,radparvar85} ($I=\pm I_1=\mp I_4$ and $I_2=I_3=0$). We will analyze the general cases when all $I_i\neq 0$. The coordinate system used in this work is indicated in Figs.~\ref{view}(a) and \ref{view}(b).

\subsection{Boundary conditions}

\begin{figure}[tb]
\centering
      \includegraphics[width=7.5cm]{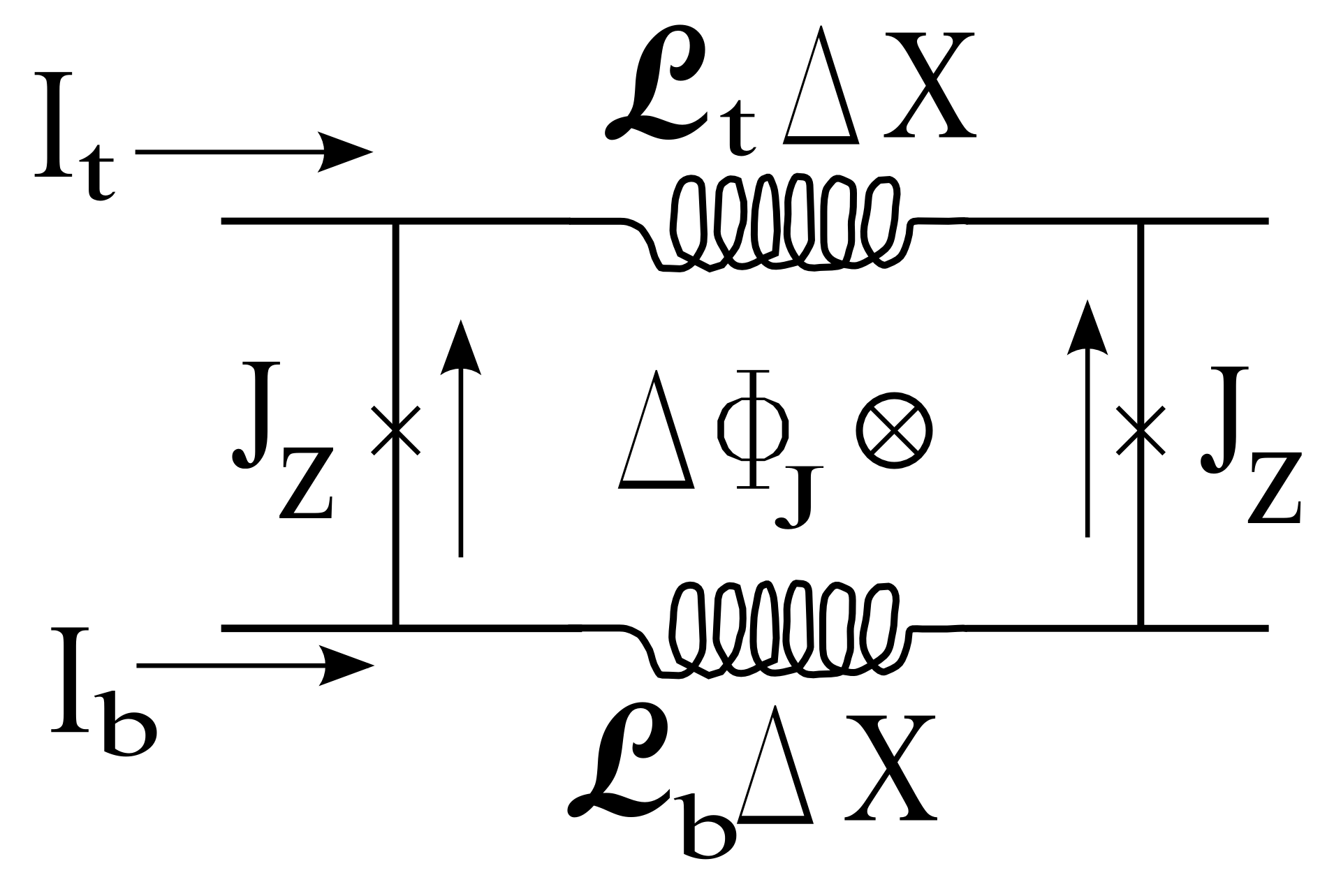}
\caption{Elementary cell of the equivalent lumped circuit for the static Josephson transmission line.\label{scheme}}
\end{figure}

Figure~\ref{scheme} shows one elementary cell of the equivalent circuit for a static Josephson junction transmission line. Classically, the two inductances have been merged in their parallel combination\cite{scott76,erne80} so that the role played by each supercurrent separately was lost; however, for our purposes it is mandatory to keep the distinction, since, in general, $\mathcal{L}_b \neq \mathcal{L}_t$. In the absence of an external in-plane field, $H_{app}=0$, the magnetic flux $\Delta \Phi_j$ linked to the cell is:
$$\Delta \Phi_j=\mathcal{L}_t \Delta X I_t -\mathcal{L}_b \Delta X I_b,$$
\noindent where\cite{mc} $\Phi_j=\Phi_0 \phi /2\pi$. Then, in the limit $\Delta X \to 0$,

\begin{equation}
\frac{\Phi_0 }{2\pi} \frac{d \phi}{dX} =\mathcal{L}_t I_t(X) -\mathcal{L}_b I_b(X).
\label{temp1}
\end{equation} 

\noindent By differentiating Eq.~(\ref{temp1}), in force of Eq.~(\ref{sG}), we end up with:

$$\mathcal{L}_t \frac{dI_t(X)}{dX} -\mathcal{L}_b \frac{dI_b(X)}{dX}=\mu_0 d_e  J_Z(X).$$
\noindent Interestingly, by integrating back the equation above over the junction length ${\rm{L}}$ and considering that, according to the notations of Fig.~\ref{view}, $I_t({\rm{L}}/2)=I_3$, $I_t({-\rm{L}}/2)=I_4$, $I_b ({\rm{L}}/2)= I_2$, and $I_b({-\rm{L}}/2)= I_1$, we find $\mathcal{L}_b+\mathcal{L}_t=\mu_0 d_e/W_{}$, i.e. $\lambda_b/W_b+\lambda_t/W_t \approx (\lambda_b+\lambda_t)/W_{}$; clearly, this can only be acceptable if $W_{} \approx W_b \approx W_t$. To overcome this limitation, we introduce a new (smaller) effective barrier penetration $d'_e\equiv W_{} \mathcal{L}_J/\mu_0 $ that takes into account the screening effect expected when the junction electrodes are wider than the tunneling barrier, namely:
\begin{equation}
d'_e = \lambda_b \frac{W_{}}{W_b} + \lambda_t \frac{W_{}}{W_t}= \frac{\lambda_b}{w_b}+ \frac{\lambda_t}{w_t},
\label{modif}
\end{equation} 

\noindent where $w_{b,t}$ denotes the width ratio $W_{b,t}/W_{}$. Being $w_{b,t} \geq 1$, then $d'_e  \leq  \lambda_b +\lambda_t \equiv d_e$. We stress that a smaller magnetic penetration results in a smaller magnetic flux through the Josephson barrier (in small junctions or at the extremities of LJTJs), but not to a reduced amplitude of the magnetic field threading the barrier. Indeed, we believe that, due to demagnetization effects in the finite-thickness films, the amplitude of the magnetic field threading the barrier is larger than that of the applied field. The effects of a reduced magnetic thickness and an increased field amplitude partially compensate each other; however, this is not a good reason to ignore them. Since the screening and demagnetization effects depends, respectively, on the film widths and thicknesses, they are independent; we leave the investigation of demagnetization in Josephson structures to a future study. In the rest of this Section we will carry out our analysis substituting $d_e$ with $d'_e$ in the magnetic Josephson equation Eq.(\ref{gra}); consequently, this new magnetic thickness also enters in expression for the Josephson penetration depth and corresponds to the $\lambda_J$ inflation, already investigated using different approaches\cite{JAP95,Ustinov,caputo}, occurring in window-type \Jos tunnel junctions. For future purposes, we also introduce the two relative inductances per unit lengths $\Lambda_b\equiv  \mathcal{L}_b/\mathcal{L}_J$ and $\Lambda_t \equiv  \mathcal{L}_t/\mathcal{L}_J=1-\Lambda_b$ and we note that, when $\mathcal{L}_b=\mathcal{L}_t$, then $\Lambda_b=\Lambda_t=1/2$ as it was implicitly assumed in all previous analytical works on LJTJs. We anticipate here that for our samples we found quite different relative inductance per unit lengths, namely $\Lambda_b \approx 0.26$ and $\Lambda_t \approx 0.74$. A practical expression for computing the bottom relative inductance not involving the junction width $W$ is:

\begin{equation}
\Lambda_b^{-1}=1 + \frac{\lambda_t W_b}{\lambda_b W_t}. 
\label{practic}
\end{equation}

Inserting Eq.(\ref{temp1}) into the magnetic Josephson equation (\ref{gra}), we obtain the local magnetic field in the barrier plane $H_Y(X)$ in terms of $I_b$ and $I_t$:

\begin{equation}
H_Y(X)= \kappa \frac{d \phi}{dX} = \frac{\Lambda_t I_t(X) -\Lambda_b I_b(X)}{W_{}}. 
\label{local}
\end{equation} 

\noindent Even when the junction electrodes are made of the same material and have the same thickness and quality, meaning that $\lambda_{t}= \lambda_{b}$, then the dependence on the electrode widths remains: $$H_Y(X)\simeq  \frac{I_t(X)W_b - I_b(X)W_t}{W_{}(W_b+W_t)}.$$ 

\noindent Equation (\ref{local}) allows us to correctly derive the boundary conditions for the static sine-Gordon equation in Eq.(\ref{sG}) in the general case $W_{} \neq W_b$ and $W_{} \neq W_t$ and in the presence of an in-plane field $H_{app}$: 

$$ \kappa{\left. \frac{d \phi}{d X} \right|_{X=-{\rm{L}}/2}}= \,  H_Y\left( -\frac{{\rm{L}}}{2}\right) = H_{app} +\frac{\Lambda_t I_4 -\Lambda_b I_1}{W_{}}, $$
\begin{eqnarray} 
\kappa{\left. \frac{d \phi}{d X} \right|_{X={\rm{L}}/2}}= \, H_Y\left( \frac{{\rm{L}}}{2}\right) =  H_{app} +\frac{\Lambda_t I_3 -\Lambda_b I_2}{W_{}}.
\label{bcnew}
\end{eqnarray}

\noindent For $\Lambda_b=\Lambda_t= 1/2$, we recover the boundary conditions by Owen and Scalapino\cite{OS} ($I_1=I_3=I$ and $I_2=I_4=0$) that were generally adopted thereafter for untrue symmetry reasons. From the boundary conditions above, it follows that: $$H_Y\left( -\frac{{\rm{L}}}{2}\right) + H_Y\left( \frac{{\rm{L}}}{2} \right) = 2 H_{app} + \frac{\Lambda_t (I_3+I_4) -\Lambda_b (I_1+I_2)}{W_{}}\, ,$$ where the last term, vanishing when $\Lambda_b=\Lambda_t$ and $I_4+I_3=I_1+I_2$, has been omitted in all previous analysis of in-line LJTJs. We like to point out that Eqs.(\ref{bcnew}) are very general and should be used to correctly describe the so-called {\it self-field effects} occurring in LJTJ. They also apply to LJTJs with mixed in-line and overlap biasing. Unfortunately, their implementation requires the separate knowledge of the bottom and top electrode inductances per unit length (rather than their sum). Indeed, the inductance per unit length was analytically derived by Chang\cite{chang} for a superconducting strip transmission line, i.e., a structure consisting of a finite-width superconducting strip over an infinite superconducting ground plane, as far as the strip linewidth $W_t$ exceeds about the insulation thickness $t_{ox}$. Definitely his results can be used when $W_b >> W_t$, but, unfortunately, no analytical expression is available when both electrodes have finite and comparable widths.

\subsubsection{Single loop}

Let us choose that the currents are positive when they flow from the left to the right junction ends, i.e., counterclockwise in the case of a ring-shaped electrode sketched in Fig.~\ref{geometry}(a). Then the boundary conditions for the single loop case can be derived as follows. With reference to Figs.\ref{view}, we recognize that for the single loop  configuration $I_1=-I_{cir}+ \alpha I$, $I_2=-I_{cir}-(1-\alpha)I$, $I_3=I$ and  $I_4=0$. Then Eqs.(\ref{bcnew}) become:

$$ \kappa{\left. \frac{d \phi}{d X} \right|_{X=-{\rm{L}}/2}}= \,  H_Y(-{\rm{L}}/2) =$$ $$=H_{app} + \frac{\Lambda_b }{W_{}}(I_{cir}-\alpha I)=H_e - \frac{\alpha \Lambda_b }{W_{}}  I;$$
\begin{eqnarray} 
\kappa{\left. \frac{d \phi}{d X} \right|_{X={\rm{L}}/2}}= \, H_Y({\rm{L}}/2) =  H_Y(-{\rm{L}}/2)+ \frac{I}{W_{}}=H_e +\left(1- \frac{\alpha \Lambda_b }{W_{}} \right)I,
\label{bcsingle}
\end{eqnarray}

\noindent with $H_e\equiv H_{app} + H_{rad}$ and 

\begin{equation}
H_{rad} = \frac{\Lambda_b I_{cir}}{ W_{}}. 
\label{Hrad}
\end{equation} 

\noindent Of course, if the loop is formed by the top, rather than the bottom, electrode, $\Lambda_t$ should replace $\Lambda_b$ in the above expression. Next, in this specific case, in order to have $I_b(-{\rm{L}}/2)=-I_{cir}+\alpha I$, $I_b({\rm{L}}/2)=-I_{cir}-(1-\alpha)I$, $I_t(-{\rm{L}}/2)=0$, and $I_t({\rm{L}}/2)=I$, it must be: 

\begin{equation}
I_b(X)=-I_{cir}+\alpha I - W_j\int_{-{\rm{L}}/2} ^{X} J_z(X')dX'
\label{ibx}
\end{equation}

\noindent and 

\begin{equation}
I_t(X) = W_j \int_{-{\rm{L}}/2}^{X} J_z(X')dX'.
\label{itx}
\end{equation}

\noindent In normalized units of $x\equiv X/\lambda_J$, the differential equation Eq.(\ref{sG}) becomes:

\begin{equation}
\frac{d^2 \phi}{d x^2} = \sin \phi(x), 
\label{ODE}
\end{equation}

\noindent with $x \in [-\ell/2,\ell/2]$ and $\ell \equiv {\rm{L}}/\lambda_J$ is the junction normalized length. Further, we will normalize the magnetic fields to $J_c \lambda_J$, so that the boundary conditions Eqs.(\ref{bcsingle}) for a LJTJ with a doubly connected base electrode are:

\begin{equation}
{\left. \frac{d \phi}{d x} \right|_{x=-\ell/2} \equiv h_l= h_e -\alpha \Lambda_b  \iota;  \quad \quad   {\left. \frac{d \phi}{d x} \right|_{x=\ell/2}} \equiv  h_r =  h_l + \iota},  
\label{bcn}
\end{equation}
\noindent where the term $\iota \equiv I/I_0=h_r-h_l$ is the external bias current $I$ normalized to $I_0 \equiv J_c W_{} \lambda_J$. With these notations, the normalized critical magnetic field $h_c$ of a short \Jos \jun is $2\pi/\ell$; further, defining $i_{cir} \equiv I_{cir}/I_0$ and $h_{rad}\equiv H_{rad}/J_c \lambda_J = \Lambda_b i_{cir}$, then $h_e \equiv h_{app}+h_{rad}= h_{app}+\Lambda_b i_{cir}$. We note that what matters now is the product $\alpha \Lambda_b$, rather than $\alpha$ itself and that the symmetry condition now corresponds to $2 \alpha \Lambda_b=1$, which can never be achieved if $\Lambda_b<1/2$. In the early eighties\cite{radparvar81}, the reported asymmetric behavior of samples that were believed to be symmetric led many experimentalists to abandon the in-line geometry in favor of the overlap one.

\subsubsection{Double loop}

We now consider the most general case of a LJTJ having both electrodes doubly connected. For the sake of simplicity, we now assume the two loops to be rectangular, as depicted in Fig.~\ref{geometry}(b) where $\beta$ is the fraction of the current $I$ diverted in the left arm of the top-electrode loop (obviously, it is impossible to realize a topologically equivalent layout with two rings). As indicated, the magnetic fluxes $\Phi_{e,t}$ and $\Phi_{e,b}$ linked, respectively, to the top and bottom loops induce the clockwise circulating currents $I_{cir,t}$ and $I_{cir,b}$ in the respective loops. Further, we recognize that $I_1=-I_{cir,b}+\alpha I$, $I_2=-I_{cir,b}- (1-\alpha)I$, $I_3=-I_{cir,t}+(1-\beta)I$ and $I_4=-I_{cir,t}-\beta I$. From Eqs.(\ref{bcnew}) the boundary conditions are:

$$\kappa{\left. \frac{d \phi}{d X} \right|_{X=-{\rm{L}}/2}}= \, H_Y(-{\rm{L}}/2) =$$ 

$$= H_{app} + \frac{\mathcal{L}_t}{\mathcal{L}_J W_{}}(I_{cir,t}-\beta I) + \frac{\mathcal{L}_b }{\mathcal{L}_J W_{}}(I_{cir,b}-\alpha I)=$$

$$= H_e - \frac{\beta \Lambda_t +\alpha \Lambda_b }{W_{}}I; $$
\begin{eqnarray} 
\kappa{\left. \frac{d \phi}{d X} \right|_{X={\rm{L}}/2}}= \, H_Y({\rm{L}}/2) =  H_Y(-{\rm{L}}/2)+ \frac{I}{W_{}},
\label{bcdouble}
\end{eqnarray}

\noindent with $H_e\equiv H_{app}+H_{rad}$ and $H_{rad}\equiv (\Lambda_b I_{cir,b}+ \Lambda_t I_{cir,t})/ W_{}$: we note that the two circulating currents interfere constructively since they flow in opposite directions, but also on the opposite sides of the tunnel barrier. The single loop configuration can be considered as a particular case of the double loop configuration in which $\beta=0$ (or $1$). Similarly the free \jun can be recovered by setting both $\alpha$ and $\beta$ in Eq.(\ref{bcdouble}) to any of their extreme values. In the rest of the paper we will limit our analysis to devices with the single loop configuration for which experimental data are available.

\subsection{Approximate solutions for LJTJs}

With the assumption that the \Jos \jun is so long that the magnetic field in its center can be neglected, $\phi_x(0)\simeq 0$, an approximate solution of Eq.(\ref{ODE}) is given by the superposition of two static non-interacting fractional fluxons pinned at the junctions extremities\cite{footnote}:

\begin{equation}
\phi(x)= \phi^l(x)+\phi^r(x),
\label{phi}
\end{equation}

\noindent with $\phi^l(x)= 4\, \textrm{sgn}h_l\, \tan^{-1} \exp -(x+\xi_l+ \ell/2)$ and $\phi^r(x)= 4\, \textrm{sgn}h_r\, \tan^{-1} \exp (x-\xi_r-\ell/2)$, and $\textrm{sgn}$ is the signum function. We observe that $\phi^l$ and $\phi^r$ do not overlap, $\phi^l(x) \, \phi^r(x)\simeq 0$, so that $\phi^2(x) \approx [\phi^l(x)]^2 + [\phi^r(x)]^2$ (similar arguments hold for the phase derivatives). As an example, for $\ell=10$, both  $\left|\phi^l(x) \, \phi^r(x) \right|$ and $\left|\phi^l_x(x) \, \phi^r_x(x) \right|$ are everywhere less than $8 \times 10^{-4}$. The phase profile in Eq.(\ref{phi}) can also be cast in the form\cite{ferrel}: $$\sin \frac{\phi(x)}{2} = \textrm{sgn}h_l\, \textrm{sech}\left( x+\xi_l+\frac{\ell}{2}\right) + \textrm{sgn}h_r\, \textrm{sech}\left(x+\xi_r-\frac{\ell}{2}\right).$$ From the phase derivative:

\begin{equation}
\phi_x(x)=  -2\,\textrm{sgn}h_l\, \textrm{sech}\left( x+\xi_l+\frac{\ell}{2}\right) +2\, \textrm{sgn}h_r\, \textrm{sech}\left( x-\xi_r-\frac{\ell}{2}\right) , 
\nonumber
\end{equation}

\noindent we infer that $\xi_l$ and $\xi_r$ are two non-negative independent constants set by the magnetic field at the boundaries $h_{r,l}$: 

\begin{equation}
h_{r,l} \equiv \phi_x\left( \pm \frac{\ell}{2}\right) = \pm2\, \sin \frac{\phi(\pm \ell/2)}{2} = 2\,\textrm{sgn}h_{r,l}\, \textrm{sech} \xi_{r,l} ,
\label{hrl}
\end{equation} 

\noindent i.e., $\xi_{r,l}=\cosh^{-1} \left|2/h_{r,l}\right|$. This indicates that, in the Meissner state, the largest possible amplitudes of the boundary fields are\cite{likharev} $h_r=h_l=2$, corresponding to $\iota=0$; then the junction critical field is $h_c=2$, corresponding to $H_c=2 J_c \lambda_J$. 

\noindent It is worth to remark that the solution in Eq.(\ref{phi}) only depends on the specific boundary conditions imposed by the system geometry. However, in the limit of vanishingly small bias current, the self-field effects disappear and the junction geometrical configuration does not affect the phase profile; in other words, for $h_e=\pm 2$, in-line, overlap and $\delta$-biased\cite{PRB10} LJTJs all have the phase profile given by Eq.(\ref{phi}) with $h_r=h_l=\pm 2$.

\begin{figure}[tb]
\centering
\subfigure[ ]{\includegraphics[width=4cm]{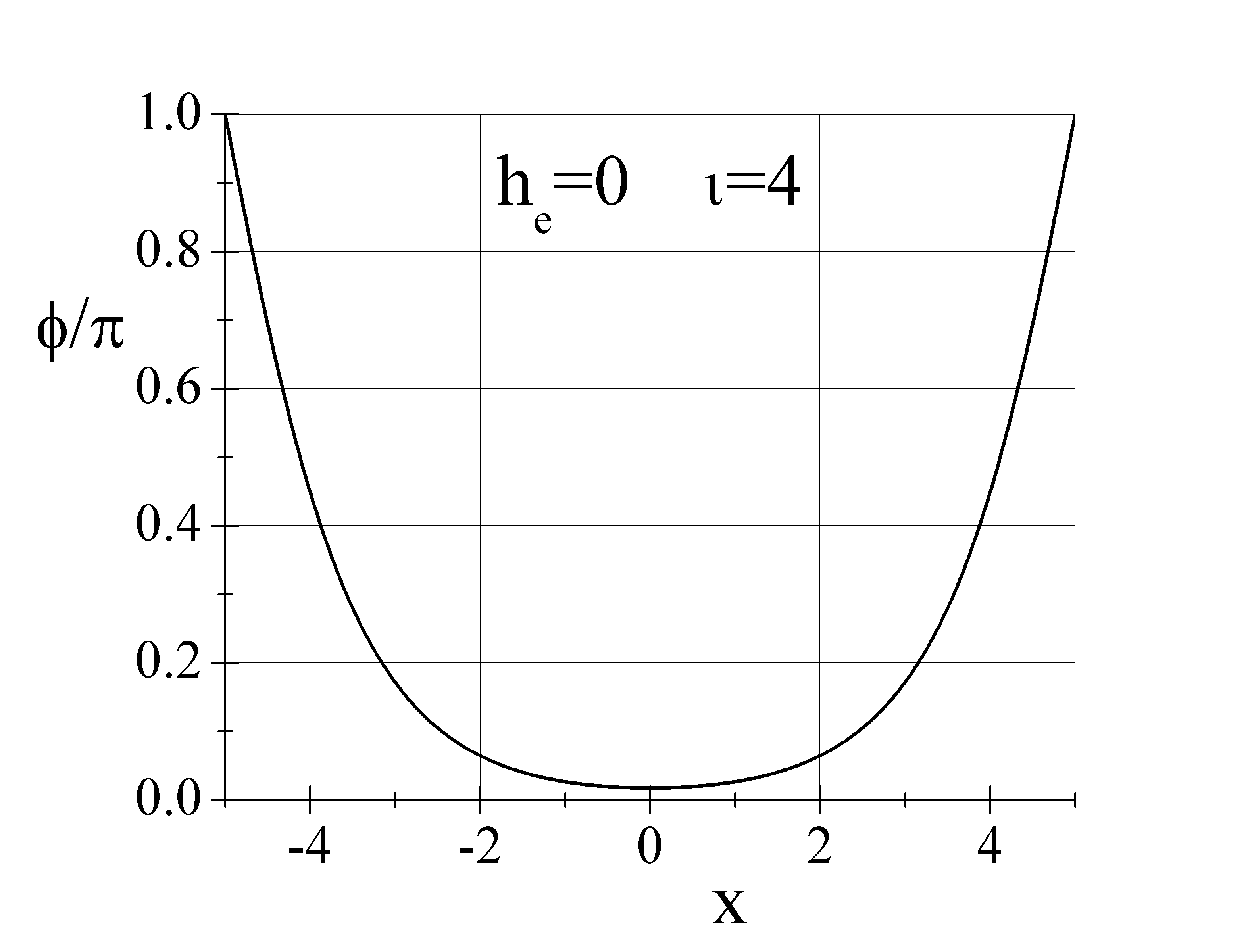}}
\subfigure[ ]{\includegraphics[width=4cm]{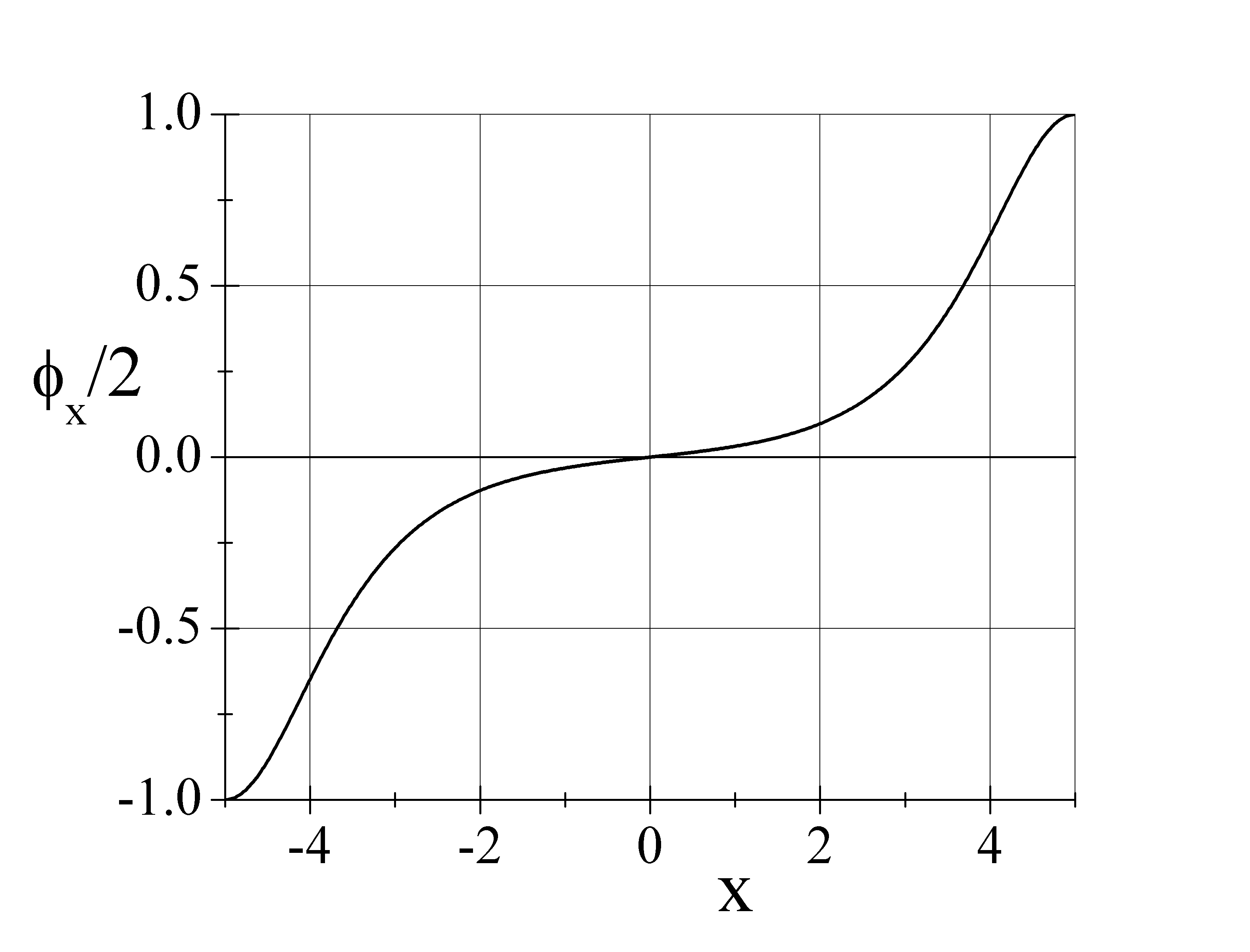}}
\subfigure[ ]{\includegraphics[width=4cm]{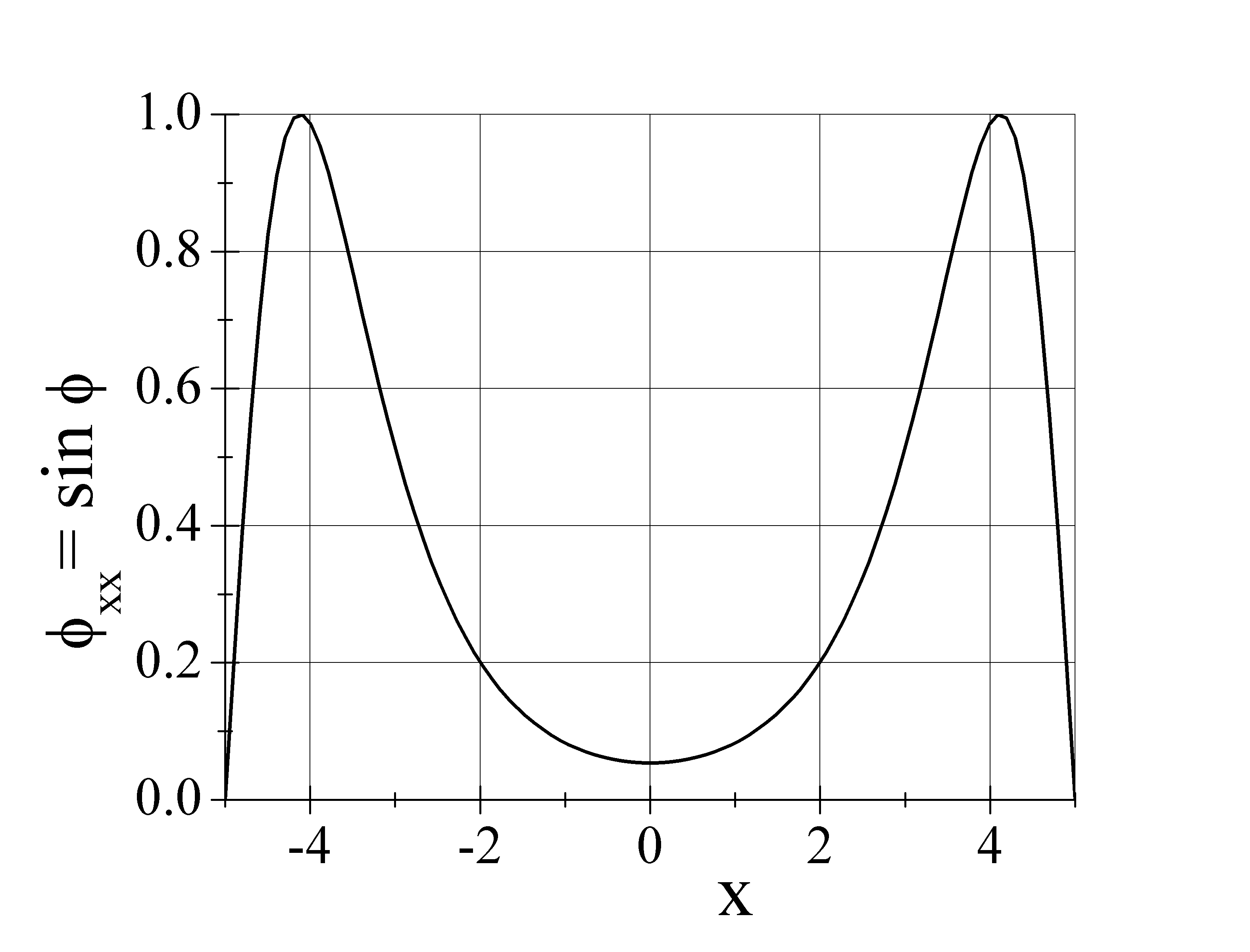}}
\subfigure[ ]{\includegraphics[width=4cm]{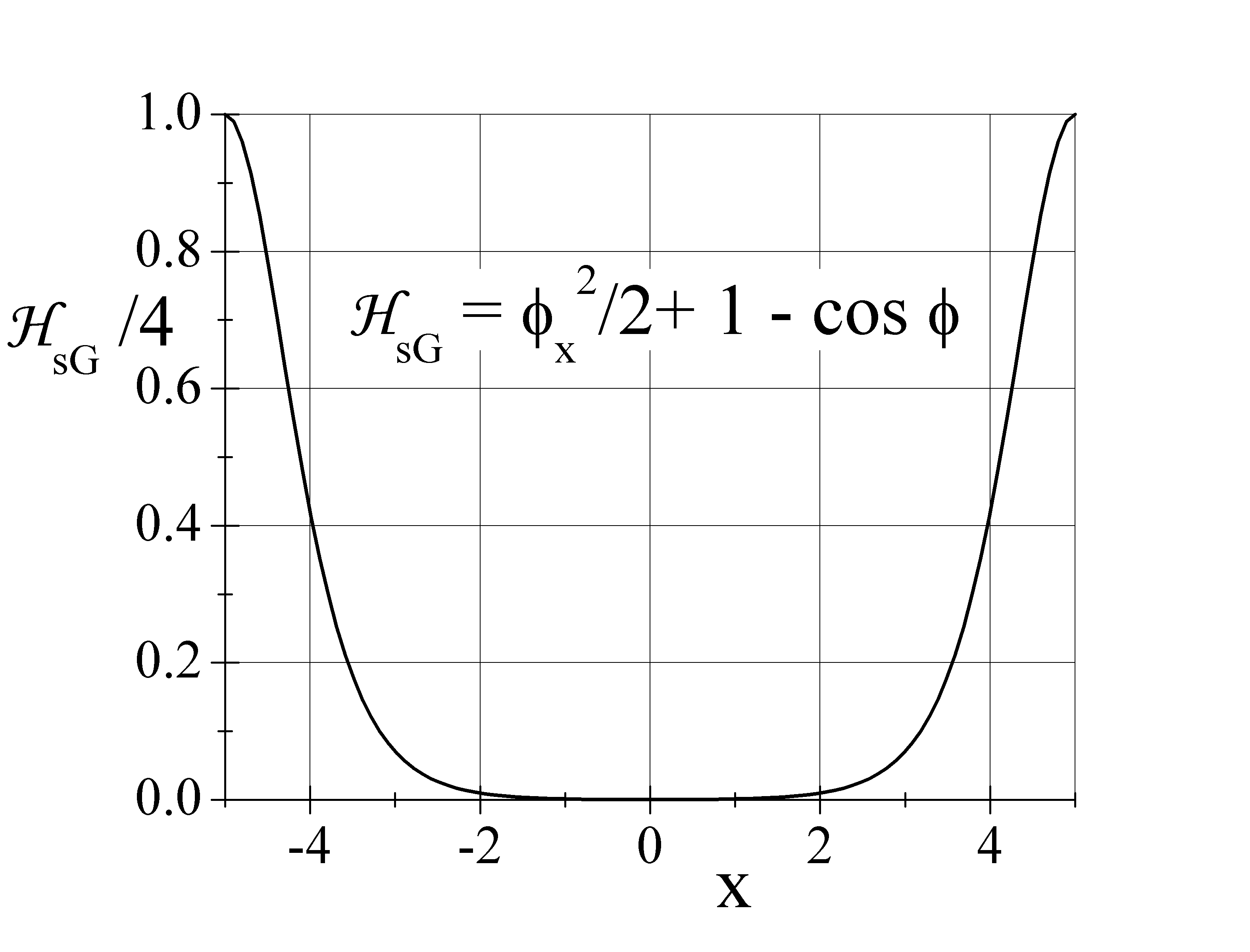}}
\caption{(a) Phase profile $\phi(x)$ of a long ($\ell=10$) in-line symmetric ($\alpha=1$) Josephson tunnel junction as in Eq.(\ref{phi}) for $h_r=-h_l=2$ that is $\xi_{l,r}=0$; (b) its first derivative $\phi_x$, (c) its second derivative $\phi_{xx}$, and (d) its normalized energy density $\mathcal{H}_{sG}$.\label{figPhi}}
\end{figure}

It can be easily proved that $\phi(0) \simeq \phi_x(0)\simeq 0$ and that, with $\xi_{l,r}=0$, then $\phi(\pm \ell/2)= \textrm{sgn}h_{r,l}\, \pi$, meaning that Eq.(\ref{phi}) corresponds to a semifluxon ($\pi$ jump) at each junction end, as shown in Fig.~\ref{figPhi}(a) for $\ell=10$ and $h_r=-h_l=2$. For the second derivative of the phase we have: $ \phi_{xx}(x) = \sin \phi = -2\,\textrm{sgn}h_l\, \textrm{sech} (x+\xi_l+\ell/2) \tanh (x+\xi_l+\ell/2)$     $ + 2\,\textrm{sgn}h_l\, \textrm{sech} (x-\xi_r-\ell/2) \tanh (x+\xi_r-\ell/2).$

\noindent  As $|h_{l}|$ ($|h_{r}|$) exceeds $h_c$, then $\xi_{l}$ ($\xi_{r}$) becomes negative, the solution in Eq.(\ref{phi}) is no longer stable and we exit the Meissner regime; in fact, for $\xi_{l,r}<0$ the phase at the extremities grows above the threshold value $|\phi|=\pi/2$ and in a dynamic scenario one (or more) integer vortex (fluxon, antifluxon) gradually develops at each extremity and moves toward the center under the effect of the Lorentz force so that some magnetic flux enters into the junction interior. 

\noindent Furthermore, for generic non-negative $\xi_{r,l}$ values, the phase difference across the junction length $\Delta \phi \equiv \phi(\ell/2)-\phi(-\ell/2)$ is:

\begin{equation}
\Delta \phi =2\, \sin^{-1} \frac{h_{r}}{2}+ 2\, \sin^{-1} \frac{h_{l}}{2},
\label{Dphi}
\end{equation}

\noindent and corresponds to what has been called a $k$-fractional vortex in Ref.\cite{goldobin05,vogel09}, where $k \equiv \Delta \phi /2 \pi$. Presently, semi ($\xi_{r,l}=0$) and fractional ($\xi_{r,l}>0$) vortices are receiving a great deal of attention in the context of $0-\pi$ transition Josephson junctions\cite{kirtley, lazarides,kemmler10}.

\subsection{Junction energy}

By applying a Lagrangian formalism\cite{barone,mc} to the sine-Gordon equation, it is possible to derive that the (static) Hamiltonian density $\mathcal{H}_{sG}$ of a LJTJ is:

\begin{equation}
\mathcal{H}_{sG}(x)= \frac{\phi_x^2 }{2}+1- \cos \phi,
\label{energ}
\end{equation}

\noindent in which the first term accounts for the magnetic energy stored in and between the junction electrodes and $(1-\cos \phi)$ is the Josephson energy density associated with the Cooper-pair tunneling current. 

\noindent The phase solution, Eq.(\ref{phi}), for LJTJs also satisfies the equality: $2(1-\cos \phi)= \phi_x^2$, meaning that for very long \juns the \Jos energy density equals the magnetic energy density (this is not true for intermediate length \juns with $\ell<2\pi$). Inserting the expression of the phase profile Eq.(\ref{phi}) in Eq.(\ref{energ}), then the junction energy density reduces to:

\begin{equation}
\mathcal{H}_{sG}(x)= [\phi_x(x)]^2= 4\, \textrm{sech}^2\left( x+\xi_l+\frac{\ell}{2}\right) + 4\, \textrm{sech}^2\left( x+\xi_r-\frac{\ell}{2} \right) ,
\label{energy density} 
\end{equation}

\noindent which is shown in Fig.~\ref{figPhi}(d) for $\ell=10$ and $\xi_{l,r}=0$. The junction energy $\hat{H}_{sG}$ can be computed from Eq.(\ref{energy density}) as a function of the boundary conditions $h_{l,r}<0$, i.e., of the external magnetic field $h_e$ and bias current $\iota$,:

\begin{equation}
\hat{H}_{sG}=\int_{-\ell/2}^{\ell/2} \mathcal{H}(x)dx = \hat{E}_0 \left( 1-\frac{\tanh \xi_l}{2}- \frac{\tanh\xi_r}{2} \right) = \hat{E}_0 \left( 1- \frac{\sqrt{4-h_r^2}}{4}-\frac{\sqrt{4-h_l^2}}{4} \right),
\label{energy}
\end{equation}

\noindent where $\tanh \xi_{l,r} = \sqrt{1-\textrm{sech}^2(\xi_{l,r})} =\sqrt{1-(h_{l,r}/2)^2}$ and $\hat{E}_0=8$ is the well-known fluxon rest mass\cite{mc} normalized to $E_0=I_0 \Phi_0/2\pi$ that, depending on the junction's electrical and geometrical parameters, represents its characteristic energy unit; typically $E_0$ is in the $10^{-18}$J range, that is, several orders of magnitude larger than the thermal energy at cryogenic temperatures. In real units, the junction energy is $E_J \equiv E_0 \hat{H}_{sG}$.

\begin{figure}[tb]
\centering
\includegraphics[width=7.5cm]{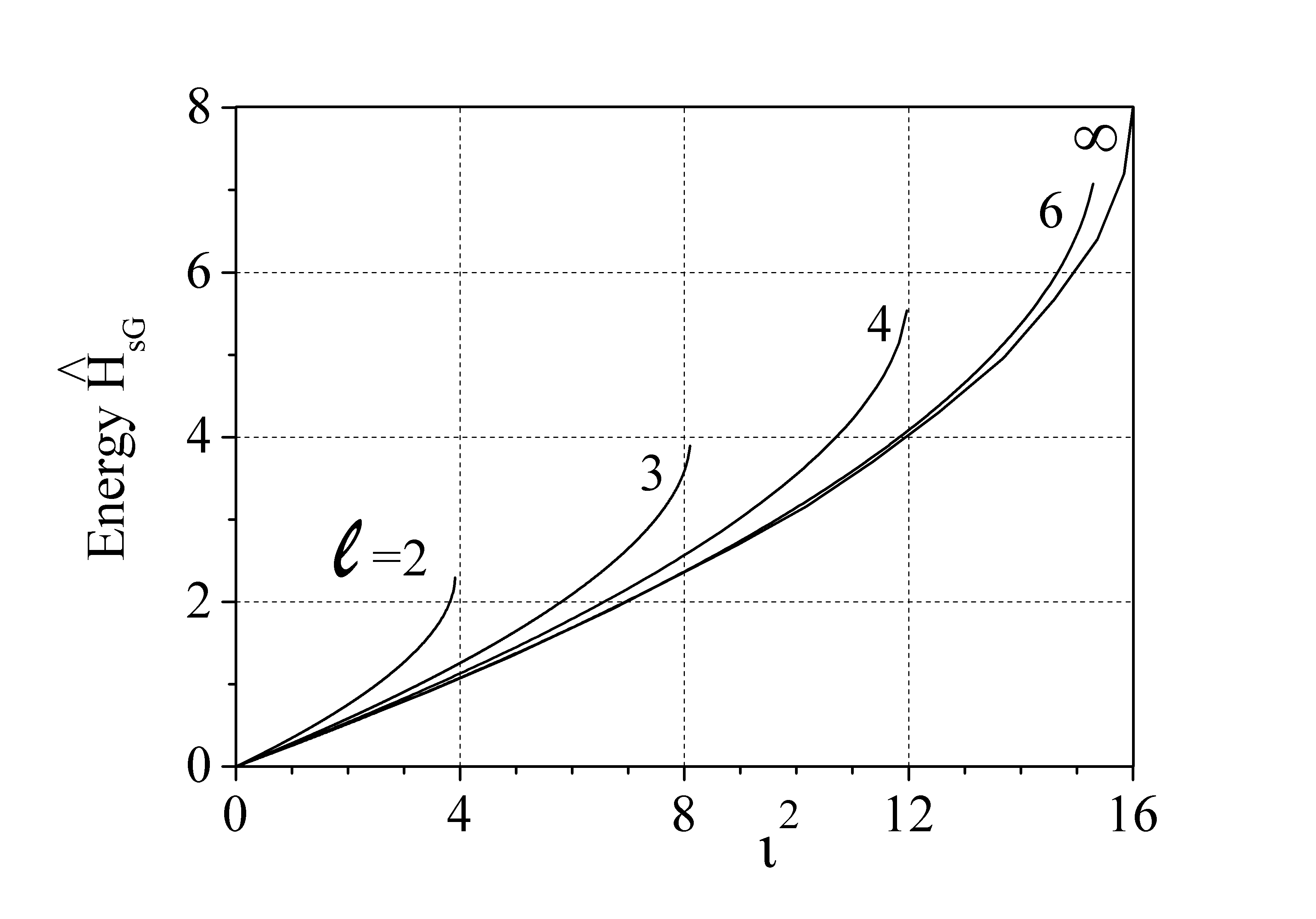}
\caption{ Numerically computed zero-field normalized junction energy $\hat{H}_{sG}$ vs. $\iota^2$ of a symmetric ($2\alpha\Lambda_b=1$) in-line Josephson tunnel junctions with  for different normalized lengths $\ell$. For $\ell \to \infty$ the analytical expression in  Eq.(\ref{EE}) was plotted. $\iota\equiv I/I_0$ is the normalized bias current.\label{energyplot}}
\end{figure}

\noindent For the particular case of a symmetric junction ($2\alpha\Lambda_b=1$) in zero field ($h_e=0$), then $h_r=-h_l=\iota/2$ so that Eq.(\ref{energy}) reduces to:

$$\hat{H}_{sG}(\iota)=\hat{E}_0(1-\sqrt{1-\iota^2/16})= $$

\begin{equation}
= 4 (\iota/4)^2 + (\iota/4)^4+ 0.5(\iota/4)^6 + O(\iota^8).
\label{EE}
\end{equation}

\noindent  Figure~\ref{energyplot} displays the numerically computed zero-field energy $\hat{H}_{sG}$ versus $\iota^2$ for symmetric in-line junctions having different normalized lengths $\ell$; for $\ell > 2\pi$ the numerical findings are very well approximated by Eq.(\ref{EE}).  As the expansion in the r.h.s. of Eq.(\ref{EE}) indicates, a LJTJ can be thought of as a non-linear inductance $2E_J/I^2$ which, in contrast to the small junction case, does not diverge at the critical current (this is because $I_c < J_c W_{} {\rm{L}}$). The largest inductance value, achieved when the junction is biased at the critical current $I_c=I_{c,max}=4I_0$, is $L_0\equiv 2 H_{sG}(I_{c,max})/I_{c,max}^2=E_0/I_0^2= \Phi_0/2\pi I_0 = \mathcal{L}_J \lambda_J$ and is the junction natural unit (typically a fraction of a picohenry) that will be used later on to normalize inductances. Normalizing the magnetic fluxes to the magnetic flux quantum $\Phi_0$, then the normalized circulating current can be written as $i_{cir}=2\pi(n-\phi_e)/l_{loop}$, where $\phi_e \equiv \Phi_e/\Phi_0$ and $l_{loop}\equiv L_{loop}/L_0$. With such notations, $l_b\equiv L_b/L_0=\Lambda_b \ell$.

\noindent Neglecting the mutual inductance effects, the system total energy $E_{tot}$ consists of two independent contributions $E_{tot}=E_{m}+H_{sG}$; the former is the magneto-static energy, $E_{m}$, stored in the inductances $L_1$ and $L_2$: $2E_{m}= L_1 I_1^2 +  L_2 I_2^2 $; the latter is the previously discussed junction energy, $H_{sG}$, which, as said before, takes into account both \Jos energies and the magnetic energy associated with the bias current $I$ flowing in the junction electrodes. In terms of normalized quantities, considering that $I_1=\alpha I - I_{cir}$ and $I_2=-(1-\alpha)I-I_{cir}$, we have that $\hat{E}_{m} \equiv E_{m}/E_0$ is: 
$$\hat{E}_{m} =  \left[ \alpha^2  l_1 + (1-\alpha)^2 l_2 \right] \frac{\iota^2}{2} + $$

\begin{equation}
+ \left[ (1-\alpha) l_2 - \alpha l_1 \right] \iota i_{cir} + \left( l_1+l_2 \right) \frac{i_{cir}^2}{2}.
\label{Em}
\end{equation}

\noindent $\hat{E}_{m}$ is minimum for $\alpha=\alpha_{min}$, where:

$$\alpha_{min}= \frac{l_2}{l_1+l_2} + \frac{i_{cir}}{\iota}=\frac{l_2}{l_{loop}-l_b} + \frac{i_{cir}}{\iota},$$

\noindent with $\iota \neq 0$. In passing we observe that, in the absence of circulating currents, $\alpha_{min}$ is independent of $\iota$. Moreover, in many cases of practical interest, $\hat{E}_{m}>>8 \geq \hat{H}_{sG}$ so that the junction energy can be neglected.

\subsection{Magnetic diffraction patterns}

Setting $h_l$ and $h_r$ at their extreme values $\pm2$ in Eqs.(\ref{bcn}), we obtain the magnetic diffraction pattern (MDP) $i_c(h_e)$ in the Meissner regime as a function of the symmetry parameter $\alpha$: 

\begin{eqnarray}
i_c(h_e) &=& \frac{2+h_e}{\alpha \Lambda_b} \qquad  \textrm{for} \quad -2 \leq h_e \leq h_{max}, \nonumber\\
&=& \frac{2-h_e}{1-\alpha \Lambda_b}  \qquad   \textrm{for} \quad h_{max} \leq h_e \leq 2,
\label{mdp}
\end{eqnarray}

\begin{figure}[tb]
\centering
\subfigure[ ]{\includegraphics[width=5cm]{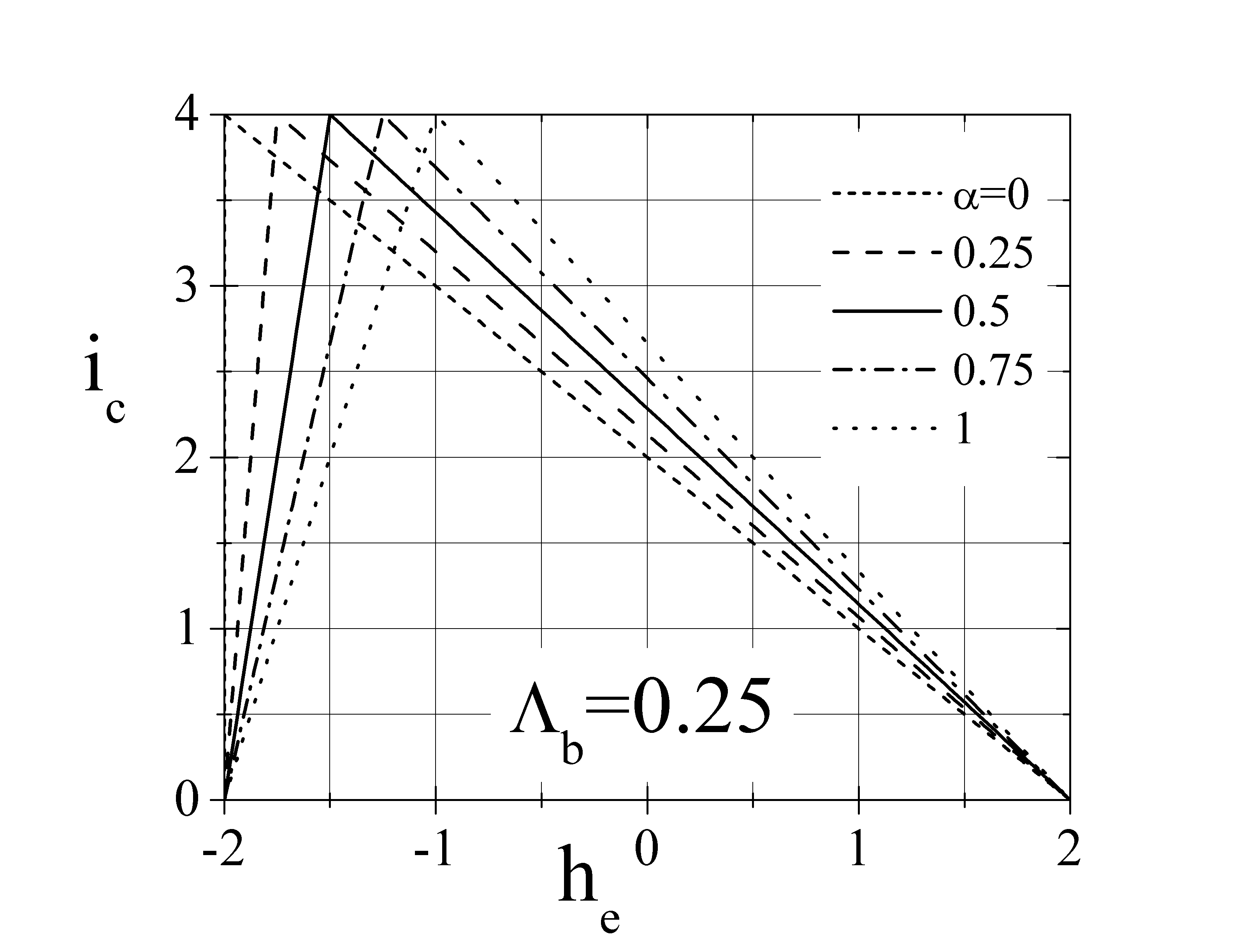}}
\subfigure[ ]{\includegraphics[width=5cm]{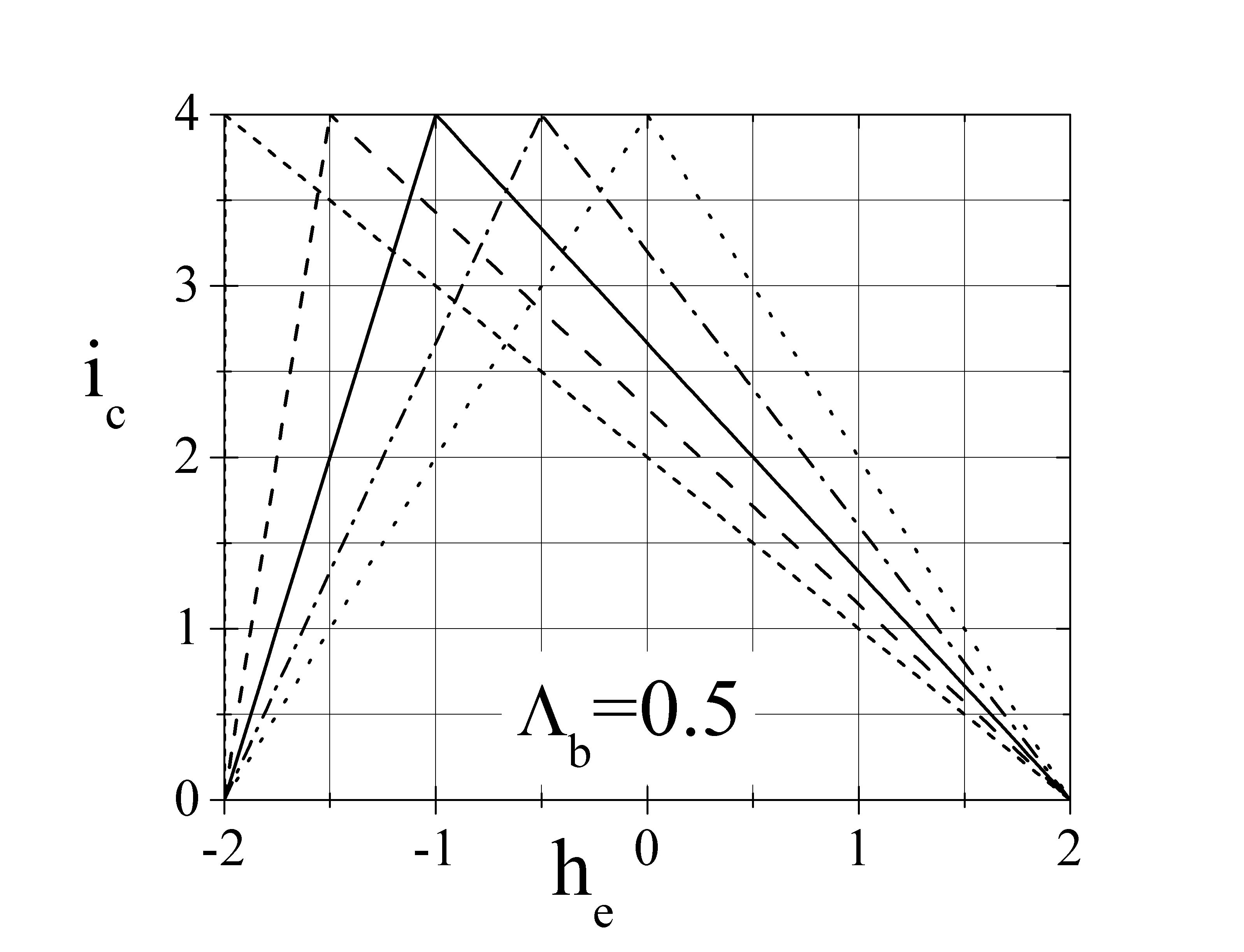}}
\subfigure[ ]{\includegraphics[width=5cm]{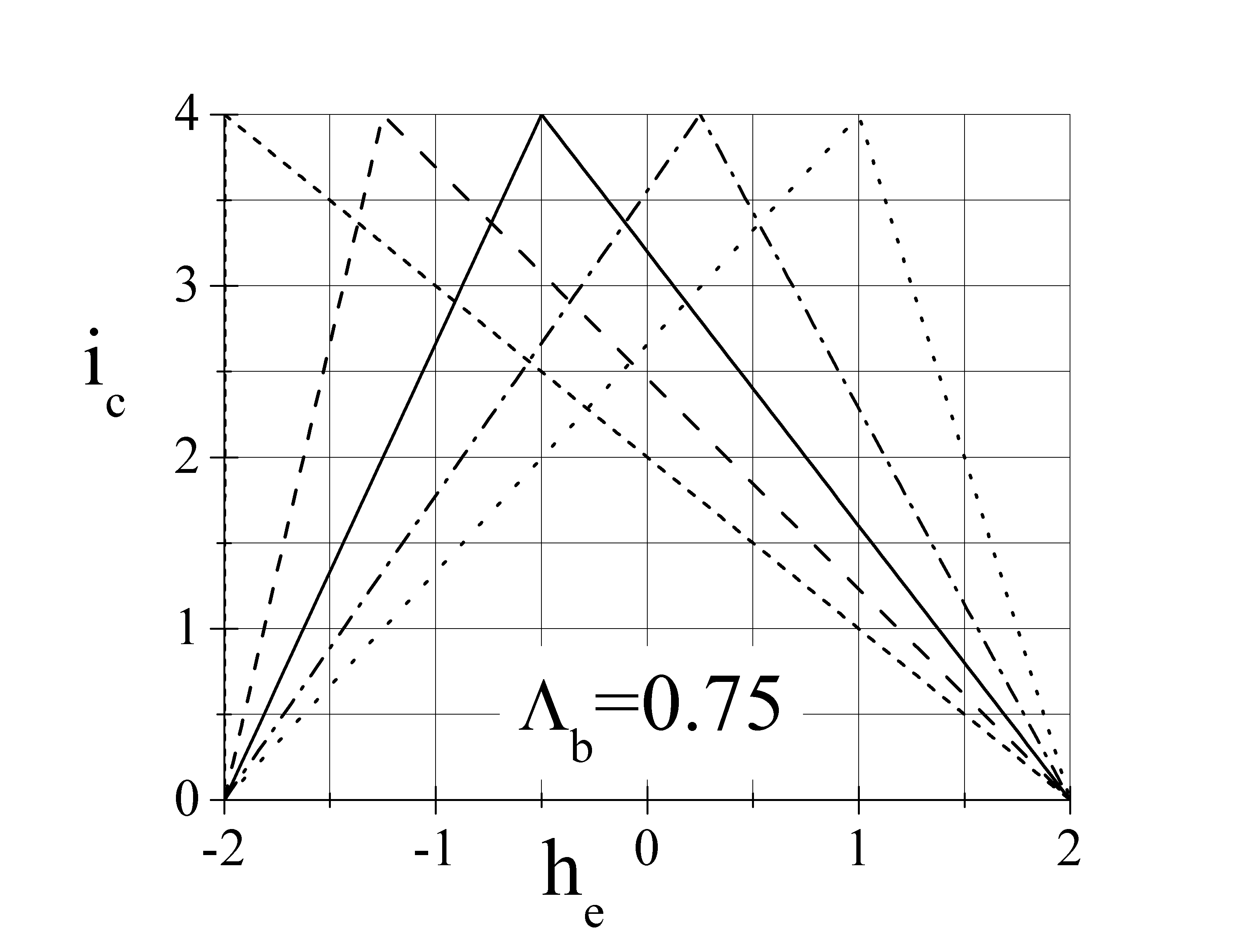}}

\caption{Magnetic diffraction patterns $i_c(h_{e})$ for a very long in-line Josephson junction with different values of the symmetry parameter $0 \leq \alpha \leq 1$ and for three $\Lambda_b$ values, namely, $0.25$, $0.5$ and $0.75$. The critical current is normalized to $J_c W_{} \lambda_J$ and the magnetic field to $J_c \lambda_J$.\label{mdpp}} 
\end{figure}

\noindent with $h_{max}\equiv 2(2\alpha \Lambda_b -1)$ being the field value which, for a given $\alpha$, yields the maximum critical current $i_c(h_{max})=4$. This $i_c$ value can be achieved only when, as depicted in Figs.~\ref{figPhi}, $h_r=-h_l=2$. The $h_{max}$ expression turns out to be very useful in the experiments to determine the product $\alpha \Lambda_b$ from the analysis of the junction MDP:

\begin{equation}
2 \alpha \Lambda_b= 1 \pm \frac{h_{max}}{2}= 1 \pm \frac{H_{max}}{H_c},
\label{hmax}
\end{equation}

\noindent in which the plus sign has to be chosen when $h_{max}$ is negative ($2 \alpha \Lambda_b < 1$) and vice versa. Figs.~\ref{mdpp}(a-c) display the MDPs for different values of the symmetry parameter $0 \leq \alpha \leq 1$ and for three $\Lambda_b$ values, namely, $0.25$, $0.5$ and $0.75$. As already stated, in general, $h_e$ results from the sum of two contributions, namely $h_{app}$ and $h_{rad}$. However, here we are neglecting the contribution to $\phi_e$ (and hence to $h_{rad}$) deriving from the current $I_t$ running in the top junction electrode; in the experiments, when needed, this contribution can be compensated by an external field perpendicular to the loop plane.

\noindent We remark that the $i_c(h_e)$ patterns are piecewise linear and, in general, they have two quite different absolute slopes $|di_c/dh_e|$ on the left and right branches. 
\noindent The wide range of linearity is very attractive for the realization of cryogenic magnetic sensors with a large dynamic range especially because large slopes can be achieved in the $[-2,h_{max}]$ interval. If the in-plane modulating field $H_e$ is the radial field $H_{rad}$ induced by a persistent current $I_{cir}$ circulating in the bottom doubly connected electrode, then $H_e=H_{rad} = \Lambda_b I_{cir}/W_{}$. It is then possible to define a current gain\cite{clarke71a,JAP11} $g_i \equiv | dI_c/dI_{cir}| = |di_c/di_{cir}|= \Lambda_b|di_c/dh_{rad}|= \Lambda_b|di_c/dh_e|$, with $i_{cir}\equiv I_{cir}/I_0$ and $h_{rad}\equiv H_{rad}/J_c \lambda_J=\Lambda_b i_{cir}$; that is:

\begin{eqnarray}
g_i &=& \frac{1}{\alpha}  \qquad \quad  \textrm{for} \,\, -2 \leq h_e \leq h_{max} \nonumber\\
&=& \frac{\Lambda_b}{1-\alpha \Lambda_b}  \qquad \, \textrm{for} \,\, h_{max} \leq h_e \leq 2.
\label{gain}
\end{eqnarray}

\noindent In zero external field, varying $\alpha$ in the range $[0,1]$, we have current gains in the range $[\Lambda_b,\Lambda_b/\Lambda_t]$. Since in all real samples $W_t<W_b$, in the case of electrodes having the same effective penetration depth $\lambda$, then $\Lambda_t>\Lambda_b$, suggesting that, for a given $\alpha$, it is preferable to realize the loop with the top electrode, rather than with the bottom one; this can also be inferred by comparing Figs.~\ref{mdpp} (a) and (c). Further, according to Eq.(\ref{gain}), large current gains can be achieved with small-$\alpha$ samples by flux biasing the loop to have $h_e < h_{max}$.  

\noindent For each trapped fluxoid the currents circulating around the loop change by an amount $\Delta I_{cir}=\Phi_0/L_{loop}$ corresponding to a jump in the critical current: 

\begin{equation}
\Delta I_{c} = g_i \Delta I_{cir} = \frac{g_i \Phi_0}{L_{loop}}.
\label{Delta}
\end{equation}

\noindent Each trapped flux quantum results in a small but detectable change in the junction critical current $I_c$. It becomes therefore possible to readout the number $n$ of  flux quanta trapped in a superconducting loop by means of an in-line LJTJ. It is worth to stress that the presented findings constitute an improvement in the state-of-the-art of current or magnetic sensors\cite{PRL12}.
  
\subsection{Fluxoid quantization}

The internal magnetic flux $\Phi_i$ within the loop is the sum of externally applied  flux $\Phi_e$ and the self-flux, $\Phi_s\equiv L_{loop}I_{cir}$, produced by the shielding current, $I_{cir}$, which circulates around the loop to restore the initial flux: 

$$ \Phi_i=\Phi_e + L_{loop}I_{cir}.$$ 

\noindent Also $\Phi_e$ results from the sum of two contributions:

\begin{equation}
\Phi_e = \Phi_a + \Phi_t = \mu_0 H_Z A_{eff} + \mathcal{L}_t \int_{-{\rm{L}}/2}^{{\rm{L}}/2} I_t(X)dX.
\label{Phie}
\end{equation} 

\noindent The first term is the applied or geometrical flux, $\Phi_a$, due to the uniform magnetic field $H_Z$ externally applied in the direction perpendicular to the loop plane and $A_{eff}$ is the effective flux capture area of the loop. For narrow loops, the pick-up areas can be well approximated by their inner areas. The second term, $\Phi_t$, is the non-linear flux contribution due to the non-uniform current $I_t$ flowing in the junction top electrode. Further, the reaction flux, $\Phi_s$, can be also expressed in terms of the currents in three inductive paths of the loop, $L_1$, $L_2$ and $L_b$:

$$\Phi_s= L_1 I_1+ L_2 I_2 + \Phi_b =$$
\begin{equation}
= (L_1+L_2)(\alpha I -I_{cir}) - L_2I + \mathcal{L}_b \int_{-{\rm{L}}/2}^{{\rm{L}}/2} I_b(X)dX 
\label{Phis}
\end{equation} 

\noindent where $\Phi_b$ is the non-linear flux contribution due to the currents $I_t$  flowing in the junction bottom electrodes. 

\noindent In the presence of an external magnetic field, $H_{app}$, applied in the loop plane and perpendicular to the long junction dimension, ${\rm{L}}$, by using Eqs.(\ref{ibx}) and (\ref{itx}), after some simple algebra we end up with:

\begin{equation}
\frac{\Phi_b}{\Lambda_b}= \mathcal{L}_t {\rm{L}} \left( \alpha I -I_{cir} \right) - \Phi_0 \frac{\Delta \phi}{2\pi} + \Phi'
\label{Phib}
\end{equation}

\begin{equation}
\frac{\Phi_t}{\Lambda_t}= \mathcal{L}_b {\rm{L}} \left( \alpha I -I_{cir} \right) + \Phi_0\frac{\Delta \phi}{2\pi} - \Phi'
\label{Phit}
\end{equation} 

\noindent with, as before, $\Delta \phi \equiv  \int_{-{\rm{L}}/2}^{{\rm{L}}/2}  \phi_X(X)dX=\phi({\rm{L}}/2)-\phi(-{\rm{L}}/2)$ being the \Jos phase difference across the junction and $\Phi'\equiv \mu_0 d_e {\rm{L}} H_{app}$ a factitious flux threading the LJTJ barrier.

\noindent The single-valuedness of the phase of the superconducting wave function around the loop (fluxoid quantization) requires\cite{london}:

\begin{equation}
\Phi_i= \Phi_a + (L_1+L_2)(\alpha I-I_{cir})- L_2 I+  \Phi_J=n \Phi_0,
\label{quantiz}
\end{equation} 

\noindent in which $n$ is an integer number, called the {\it winding} number, corresponding to the number of flux quanta trapped in the ring and $\Phi_J=\Phi_t+\Phi_b$ is the non-linear contribution to the internal flux due to the currents $I_t$ and $I_b$ flowing in the junction electrodes:

$$ \Phi_J= \mathcal{L}_t \int_{-{\rm{L}}/2}^{{\rm{L}}/2} I_t(X)dX + \mathcal{L}_b \int_{-{\rm{L}}/2}^{{\rm{L}}/2} I_b(X)dX =$$

\begin{equation}
= 2 \Lambda_t \Lambda_b {\rm{L}} \mathcal{L}_J (\alpha I -I_{cir}) + \left( \Lambda_t-\Lambda_b \right)  \left(\frac{\Phi_0}{2\pi} \Delta \phi - \Phi' \right).
\label{PhiJ}
\end{equation} 

\noindent Inserting the expression Eq.(\ref{PhiJ}) for $\Phi_J$ into Eq.(\ref{quantiz}) and switching to normalized units, the fluxoid quantization law reads:

\begin{equation}
\Delta \Lambda \Delta \phi= (l_2-\alpha l_{loop} ) \iota + l_b \Delta \Lambda ( \alpha \iota -i_{cir})+ 2\pi \Delta \Lambda \phi',
\label{quanti2}
\end{equation} 

\noindent where $\Delta \Lambda \equiv \Lambda_t - \Lambda_b$. By its definition, $\left| \Delta \Lambda \right| <1$.  If the material, quality and thickness of the junction base and top films are similar, then $\lambda_{t} \simeq \lambda_{b}$ and, consequently, the expression for $\Delta \Lambda$ reduces to $\Delta \Lambda\simeq (W_b-W_t)/(W_b+W_t)\geq 0$. If $\Lambda_t=\Lambda_t$, then, for symmetry reasons, the non-linearity of the Josephson element does not play any role, as far as concerns the fluxoid quantization and we find that, as expected, $\alpha$ is given by the ratio: 

\begin{equation}
\alpha_0 \equiv \frac{L_2}{L_{loop}}=\frac{l_2}{l_{loop}}.
\label{alpha0}
\end{equation} 

\noindent It is interesting to observe that the same result is obtained by minimizing the magnetic energy $\hat{E}_{m}$ in Eq.(\ref{Em}), if $l_1$ were replaced by $l_1+l_b= l_{loop}-l_2$. In other words, for $\Delta \Lambda=0$, both the fluxoid quantization and the energy minimization carry the same information, although they are independent principles. Eq.(\ref{quanti2}) is the reason why in our analysis we need to distinguish $\mathcal{L}_b$ from $\mathcal{L}_t$ so that, in general, the fluxoid quantization imposes a constraint on the \Jos phase difference $ \Delta \phi$ across the junction. Provided $\Delta \Lambda \neq 0$, Eq.(\ref{quanti2}) can be rearranged as: 

\begin{equation}
\Delta \phi=\Delta \Lambda^{-1}l_{loop}(\alpha_0-\alpha) \iota  + \Lambda_b \ell (\alpha \iota -i_{cir})+ 2\pi \phi',
\label{constraint}
\end{equation}

\noindent The fluxoid quantization leaves the parameter $\alpha$ as a still unknown quantity that has to be determined by using the energy minimization principle. Since the system total energy $E_{tot}$ is the sum of two contributions, it might have more than one local minimum corresponding to states with different critical current $I_c$. We remark that the fluxoid quantization and the energy minimization are independent principles and as such they must be satisfied simultaneously. Summarizing, the analysis of a LJTJ having a doubly connected electrode requires the self-consistent solution of the differential equation in Eq.(\ref{ODE}) with the boundary conditions Eq.(\ref{bcn}), the constraint Eq.(\ref{constraint}) and the further requirement of energy minimization with respect to the parameter $\alpha$ that will depend on the system parameters $\iota,h_{app},h_{rad},l_2,l_1,l_b,\Lambda_b$ and can be considered as a system degree of freedom. This complex task cannot be carried out analytically and, in general, one should use rather involved numerical methods.

\noindent The bond imposed by Eq.(\ref{constraint}) frustrates the Josephson phase $\phi(x)$ along the junction. As far as $\iota<<4$, $\phi(x)$ can adapt its profile to the constraint. However, increasing $\iota$ one reaches a point when no phase profile is compatible with the corresponding bond and a premature switching occurs. The capacity of handling the phase frustration grows with the normalized length of the junction. Ultimately, in the system under investigation, the fluxoid quantization results in a phase frustration which, in turn, reduces the junction critical current. With a fixed phase difference, dynamic processes such as the resonant fluxon motion\cite{PRB09} cannot exist any longer; nevertheless, flux flow processes will survive and their stability could even be enhanced by a fixed $\Delta \phi$.

\noindent For the double loop devices $\alpha$ and $\beta$ can be considered as two degrees of freedom for the system and we shall have to apply two fluxoid quantization rules which act as a double constraint on the phase difference $\Delta \phi$ at the junction extremities. This is equivalent to fix the phases $\phi(-\ell/2)$ and $\phi(\ell/2)$ at the junction extremities. As shown in Fig.~\ref{figPhi}(c), the peaks of the supercurrent density are not at the edges of the junction but are positioned where is needed to maximize the total current. Therefore a LJTJ can carry a net supercurrent despite the boundary phase coercions; however, in this case, only internal dynamic processes, such as the resonant plasma oscillations, will be allowed. The two quantization rules together with the energy minimization condition allow to determine both $\alpha$ and $\beta$. However, with two loops, the system total energy should also include the mutual magnetic interaction. We postpone the thorough analysis of such devices to a future work.

\subsection{Remarks}

\noindent In our model of window-type LJTJs we have neglected the thickness of the oxide layer which is correct as far as we deal with the tunneling region. A more realistic picture of real devices should consider that in the idle region surrounding the tunnel area the insulation between the bottom and top electrode is provided by an oxide layer typically made of a deposited $SiO_x$ layer and/or a anodic oxide. The total thickness of this layer is comparable or even larger than the electrode penetration depths $\lambda_{b,t}$ (and might also be comparable with the strip width). In Ref.\cite{JAP95}, each electrode was modeled as a parallel combination of two stripes having quite different oxide thicknesses resulting in a rather involved expression of the effective magnetic thickness. The expression of $d'_e$ proposed in Eq.(\ref{modif}) should therefore be considered as just a first approximation which needs to be refined. Both magneto-static simulations and/or properly devised experiments could improve our knowledge on this topic.

\noindent In this Section the consequences of the fluxoid quantization were derived for a LJTJ built on a narrow superconducting loop and the resulting phase constraint in Eq.(\ref{constraint}) also contains a term proportional to the junction normalized length $\ell$. However, it is not clear which junction length maximizes (or minimizes) the effects of phase frustration. Further, our results cannot be extrapolated to the limit of small junctions which are not affected by self-field effects, i.e., $h_r=h_l$. In previous (not published) experiments, no evidence of the fluxoid quantization was observed in the magnetic diffraction patterns of small \Jos \juns built on large inductance loops $l_b<<l_{loop}$. Nevertheless, we expect a different behavior in the limit $l_b\simeq l_{loop}$, that is, when the junction energy dominates the loop energy.

\section{EXPERIMENTS}

\subsection{Experimental setup}

\noindent Our setup consisted of a cryoprobe inserted vertically in a commercial $LHe$ dewar ($T=4.2$K). The cryoprobe was magnetically shielded by means of two concentric $Pb$ cans and a cryoperm one; in addition, the measurements were carried out in an rf-shielded room. The external magnetic field could be applied both in the chip plane or in the orthogonal direction. In fact, the chip was positioned in the center of a long superconducting cylindrical solenoid whose axis was along the $Y$-direction [see Fig.~\ref{view}(a)] to provide an in-plane magnetic field $H_{app}=H_Y$. A transverse magnetic field, $H_Z$, was applied by means of a superconducting cylindrical coil with its axis oriented along the $Z$-direction; this transverse field induces a controllable shielding current $I_{cir}$ circulating  in the superconducting loop that, in turn, generates a radial field, $H_{rad}$, in the insulating layer of the Josephson structure that algebraically adds to $H_{app}$. The field-to-current ratio was $3.9\, \mu$T/mA for the solenoid and $4.4\, \mu$T/mA for the coil. These values have been numerically obtained from Comsol Multiphysics\cite{comsol} magneto-static simulations in order to take into account the strong correction to the free-space solution due to the presence of the close fitting superconducting shield\cite{SUST09}. The effects of a transverse field on the static properties of both short\cite{JAP08} and long\cite{JAP07} Josephson tunnel junctions of various geometries have recently been investigated.

\subsection{Samples}

\begin{figure}[tb]
\centering
\subfigure[ ]{\includegraphics[width=7.5cm]{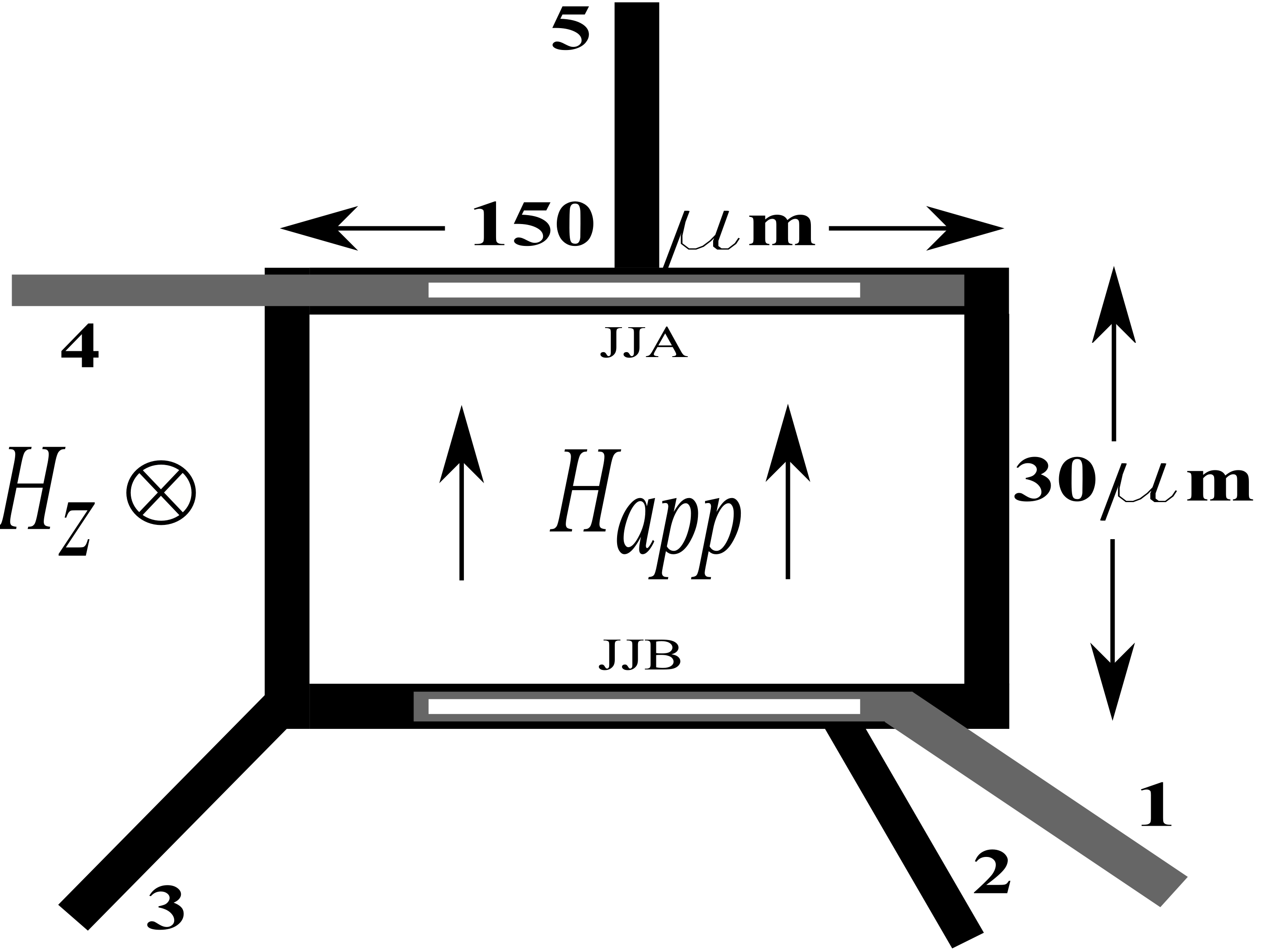}}
\subfigure[ ]{\includegraphics[width=7.5cm]{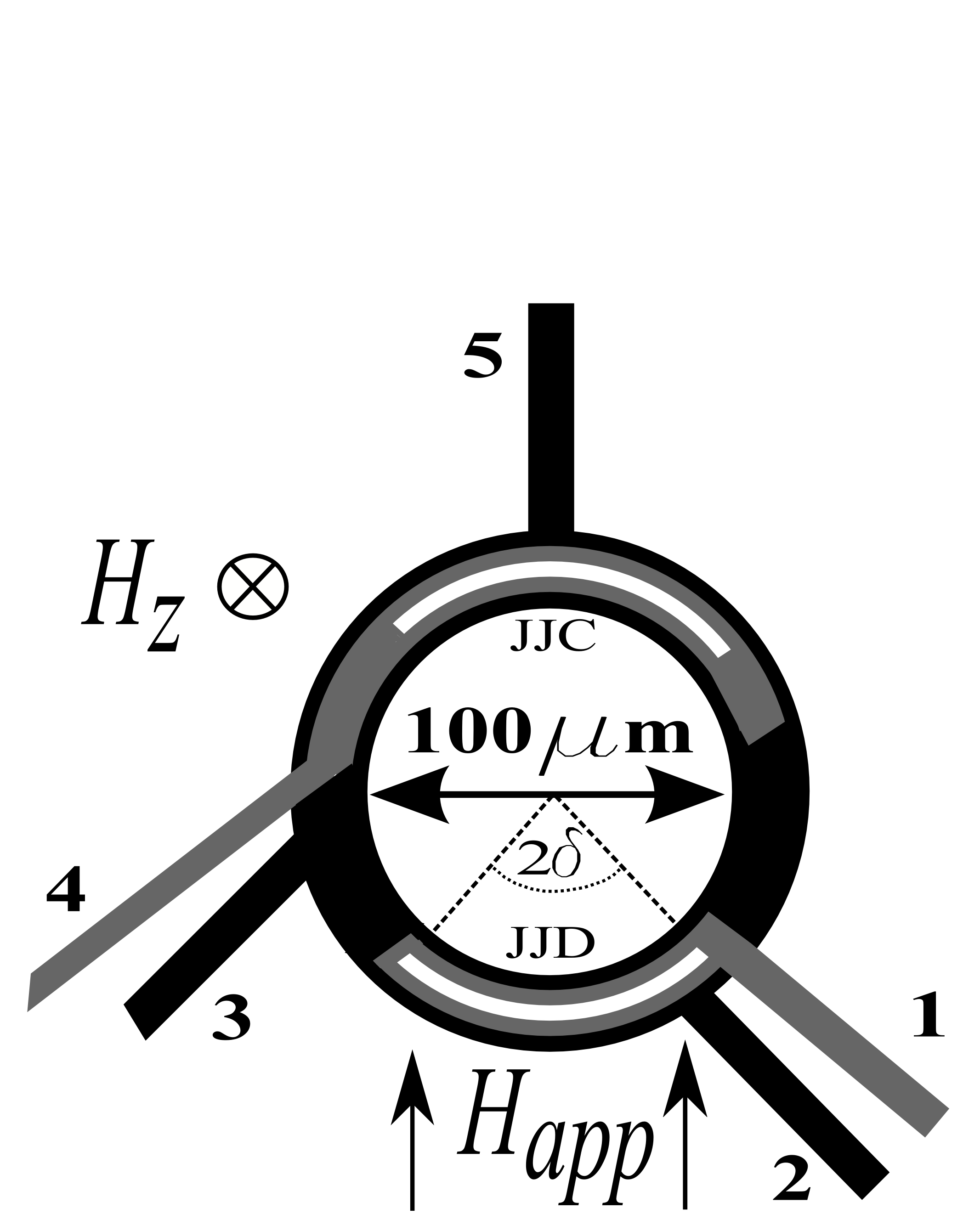}}
\caption{Geometrical layouts (not to scale) of our samples: (a) rectangular and (b) circular loop. The semi-angular length of the junction is $\delta \approx 1$.\label{HD24}}
\end{figure}

In Figs.~\ref{HD24}(a) and (b) we report the two (topologically equivalent) geometrical configurations used for our experiments; the rectangular loop in Fig.~\ref{HD24}(a) has a mean perimeter approximately equal to $2\pi$ times the mean radius $R=53\,\mu$m of the ring-shaped loop in Fig.~\ref{HD24}(b). Since the semi-angular length $\delta \equiv {\rm{L}}/2R \approx 1$, to a first approximation, the curvature of the junction built on top of the ring can be ignored. In both cases each loop allocates two in-line junctions sharing the doubly connected base electrode; they can be biased separately and under different bias configurations corresponding to different $L_2/L_{loop}$ ratios, i.e., $\alpha_0$ values. In addition, if the bias current is applied through their respective counter-electrodes (terminals $1$ and $4$ in Fig.~\ref{HD24}), the two junctions are series biased. Here, we will only present experimental data on single loop devices with just a single junction biased; the common biasing of both junctions has been intended for a different class of experiments aimed to improve the results of Ref.\cite{PRB09} and will be the subject of future work.

\noindent To correctly analyze our geometrical configurations, we have to introduce one more path inductance corresponding to the loop section acting as the base electrode for the passive LJTJ. Furthermore, if both junctions were biased simultaneously, the fluxoid quantization would involve the phase differences across both junctions and the total system energy would include the two junction's energies, as well. However, if only one of the junctions is biased, then the model developed for the single loop can be fully adapted with the caveat that properly calculated contributions have to be added to the inductive paths $L_1$ and/or $L_2$.

\noindent We evaluated the loop inductances considering the loops as isolated superconducting narrow loops:  the results are $90$ and $240$pH for the rectangular and circular loop, respectively. However, these are overestimated values, since the portions of the loop covered by the counter electrode have a lower inductance per unit length.  The experimental data will provide more accurate values.

\noindent In the experiments, we used high quality $Nb/Al-Al_{ox}/Nb$ LJTJs fabricated on silicon substrates using the trilayer technique in which the Josephson junction is realized in a window opened in a $200\,$nm thick $SiO_2$ insulator layer. We measured a large number of junctions with width $W_{}=1.5\,\mu$m and  length ${\rm{L}}=100\,\mu$m and \Jos current density $J_c\simeq 3.6\,$kA/cm$^2$, as measured in small junctions realized in the same fabrication batch. The two junctions on a given loop only differ by the longitudinal idle regions which are $50\, \mu$m for JJA and JJC and $1\, \mu$m for JJB and JJD. The nominal thicknesses and widths of the bottom and top electrodes were, respectively, $d_b=100\, $nm, $W_b=6\,\mu$m, $d_t=350\, $nm, and  $W_t=4\,\mu$m , so that, assuming\cite{valery} $\lambda_{L,Nb}=90\,$nm, it is $\lambda_b=45\,$nm, $\lambda_t=85\,$nm and the effective magnetic penetration $d'_e$, as given in Eq.(\ref{modif}), resulted to be $d'_e \approx 43\,$nm. The same values in Eq.(\ref{practic}) give $\Lambda_b=1-\Lambda_t \approx 0.26$ corresponding to $\mathcal{L}_t \approx 3 \mathcal{L}_b$ and $\Delta \Lambda\approx 0.48$. The value of $d'_e$, together with the critical current density given above, yields $\lambda_J\simeq 12.7\,\mu$m; henceforth, our junctions had a nominal normalized length ${\rm{L}} / \lambda_J \simeq 7.9$ and we can treat them as long (${\rm{L}} >2\pi \lambda_J$), one-dimensional ($W< \lambda_j$) Josephson tunnel junctions.  (By assuming that the magnetic penetration was $d_e =\lambda_b+ \lambda_t \approx 130\,$nm, we would get $\lambda_J\simeq  7.3\,\mu$m and  $\Lambda_b=\lambda_b/d_e=1-\Lambda_t \approx 0.35$.)
 
\subsection{Magnetic diffraction patterns}

\begin{figure}[b]
\centering
\subfigure[ ]{\includegraphics[width=7.5cm]{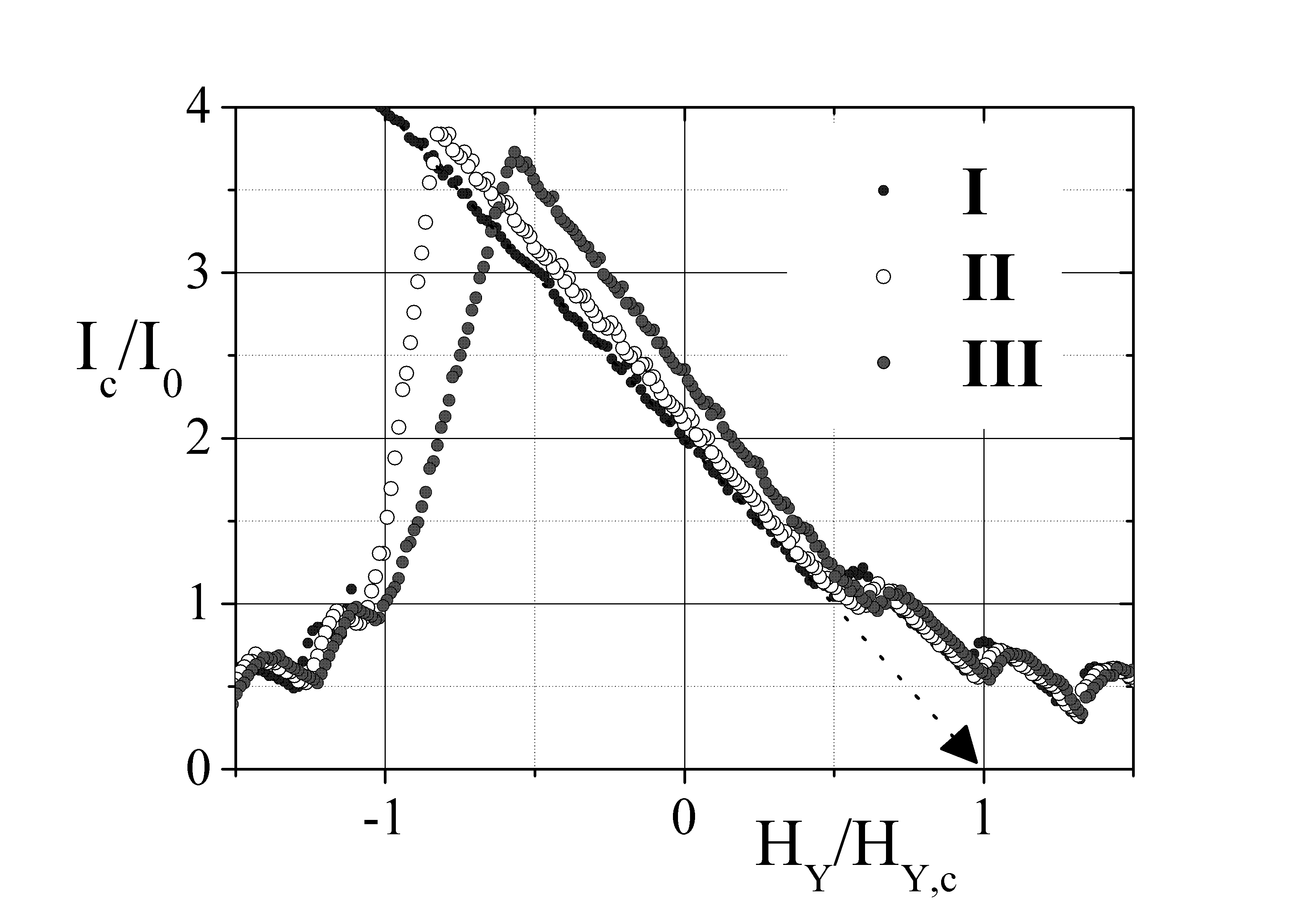}}
\subfigure[ ]{\includegraphics[width=7.5cm]{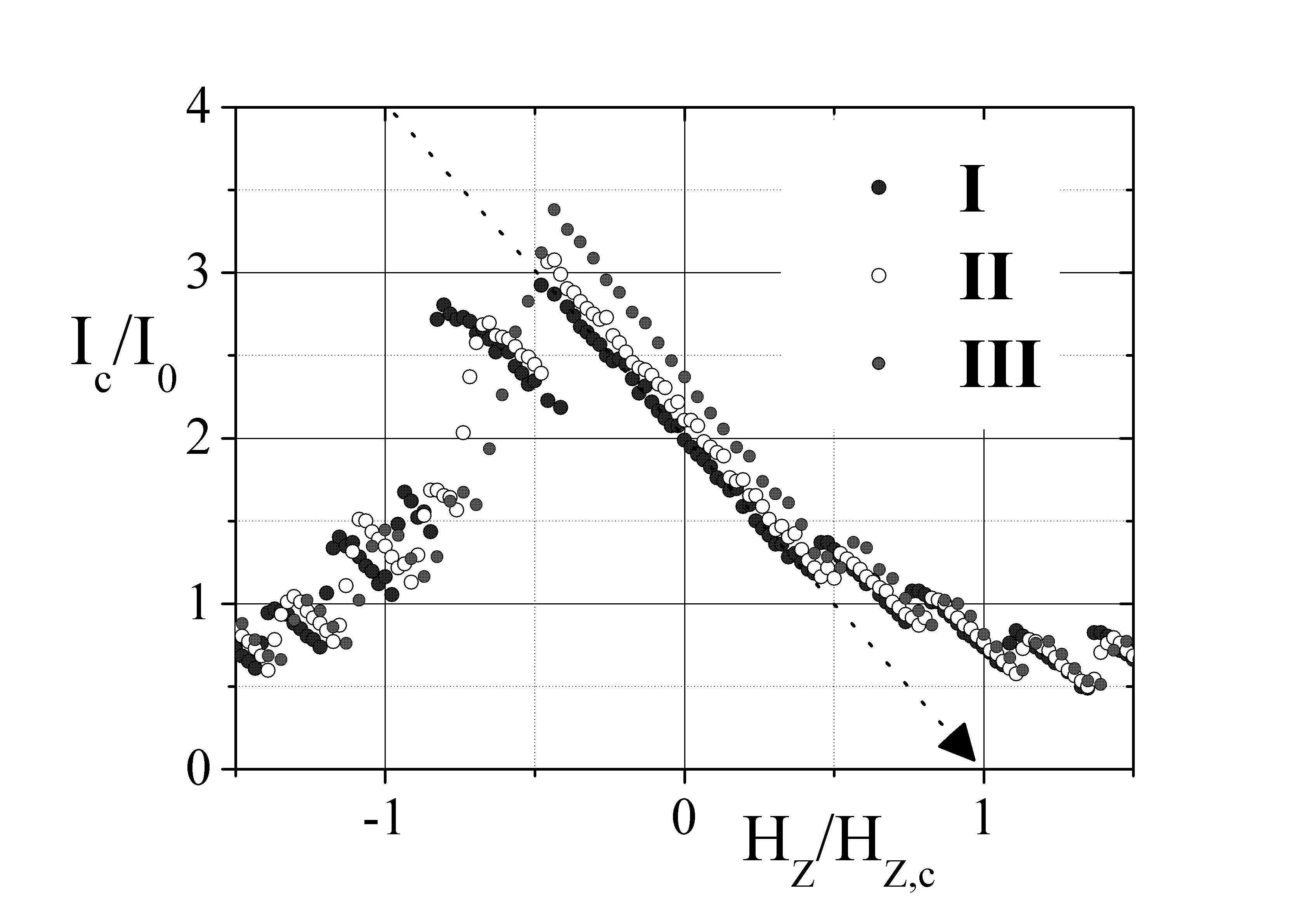}}
\caption{Experimental magnetic diffraction patterns of a long Josephson tunnel junction built on top of a rectangular loop (sample JJB) under different bias configurations and applied magnetic field orientation: (a) in-plane and (b) transverse magnetic field. With reference to Fig.~\ref{HD24}(b), the voltage is measured across the terminals $1$ and $2$, while the bias current is applied through the contacts:  $1$ and $2$ for curves I, $1$ and $3$ for curves II and $1$ and $5$ for curves III. $I_0=460\,\mu$A, $H_{Y,c}=490\,$A/m, and $H_{Z,c}=80\,$A/m at $T=4.2\,$K.\label{exp}}
\end{figure}

\noindent On real samples, the measurement of the maximum supercurrent versus the external field yields the envelop of the lobes, i.e., the current distribution switches automatically to the mode which, for a given field, carries the largest supercurrent. Sometimes, for a given applied field, multiple solutions are observed on a statistical basis by sweeping many times the \jun current-voltage characteristic. Figs.~\ref{exp}(a)-(b) display, respectively, the in-plane and transverse MDPs for a LJTJ built on top of a rectangular loop under different bias configuration (resulting in different $L_1/L_2$ ratios). Qualitatively similar results (not reported) were obtained for samples with annular geometry. In both figures, the data labels I, II and III indicate that the junction bias current was applied through, respectively, the terminals $1$ and $2$, $1$ and $3$ and $1$ and $5$. We remark that, as expected, all the MDPs are symmetric with respect to inversion of both the junction bias current, $I_c$, and the applied magnetic fields, $H_Y$ or $H_Z$. In  Figs.~\ref{exp}(a)-(b) $I_0=460\,\mu$A, $H_{Y,c}=490\,$A/m, and $H_{Z,c}=80\,$A/m. As shown by the dotted line, the critical field values were obtained by extrapolating to zero the linear branches of the MDP main lobe (by definition the critical field does not depend on the bias configuration). The theoretically expected value of $I_0=J_c W \lambda_J \simeq 690\, \mu$A exceeds its experimental counterpart by $50\%$. Since in the fully asymmetric configuration I the self-flux effects are absent (with $L_2=\alpha_0=0$, no fraction of the bias current circulates in $L_1$), at a qualitative level, we recognize this discrepancy as the signature of the phase frustration in Eq.(\ref{constraint}) provided by the fluxoid quantization. Put in a different way, the presumed harmless fact to have a doubly connected electrode depresses the critical current with about $35\%$. Unfortunately, even in absence of external fields and with the winding number $n$ set to zero, no analytical expression can be found for the value of $I_0$ expected in presence of the doubly connected electrode.

\noindent Further, the expected value of the in-plane critical field, $H_{app,c}\equiv 2 J_c \lambda_j \simeq 915\,$A/m is almost twice as large as the measured $H_{Y,c}$; this discrepancy can be ascribed to the previously mentioned in-plane demagnetization that, in the barrier proximity, {\it squeezes} the field lines of any magnetic field applied in the junction plane. Interestingly, good agreement was found instead for the transverse critical field, $H_{Z,c}$, which makes us confident about demagnetization effects for the in-plane fields. In fact, according to Eq.(\ref{Hrad}), the radial field experienced by a LJTJ built on a superconducting loop is: 

\begin{equation}
H_{rad} = \frac{\Lambda_b I_{cir}}{W_{}} \approx \frac{\mu_0 \Lambda_b A_{eff}H_Z}{W_{} L_{loop}} ,
\label{hrad}
\end{equation}

\noindent in which $L_{loop} I_{cir} \approx \mu_0 H_Z A_{eff}$ is the magnetic flux threading the loop and $A_{eff}$ the effective flux capture area of the loop. By definition also $H_{rad,c}=2 J_c \lambda_j$, then Eq.(\ref{hrad}) provides the following expression for the expected transverse critical field, $H_{Z,c}^{th}$:

$$H_{Z,c}^{th} = \frac{2 J_c \lambda_j W_{} L_{loop}}{\mu_0 \Lambda_b A_{eff}}.$$

\noindent  For narrow loops, as in our cases, the pick-up areas can be well approximated by their areas; for the rectangular loop of Fig.~\ref{exp}(a), $A_{eff} \approx 4.5 \times 10^{-3}\,$mm$^2$. By inserting the correct values in the last equation we get $H_{Z,c}^{th} \approx 84$A/m. According to Eq.(\ref{Hrad}), when $H_Z=\pm H_{Z,c}$ then the shielding current is $I_{circ}\approx \pm 2 I_0/\Lambda_b$; at variance with the applied supercurrent $I$, the circulating currents are not bonded to the $[-4I_0,4I_0]$ interval.
  
\noindent We note that the experimental in-plane MDPs in Fig.~\ref{exp}(a) closely reproduce the theoretical ones of Fig.~\ref{mdpp}(a), apart from the fact that the largest critical current values $I_{c,max}$ should be independent on the biasing terminals. However, the biasing configuration II and III are characterized by larger and larger circulating currents which depress the junction critical current even further. As already stated, our derivation of the MDPs did not take into account the self-flux effects. However, we were able to compensate these effects by a proper small transverse field $H_Z$.

\noindent According to Eq.(\ref{hmax}), the product $\alpha_0  \Lambda_b$ can be determined from the field value $H_{max}$ corresponding to the largest critical current $I_{c,max}$. We stress that the position of $H_{max}$ remains independent on any possible suppression of the critical current $I_c$. Taking the fully asymmetric configuration I as a reference, $\alpha_0=0$ when $H_{max}=H_c$, the obtained values of $\alpha_0 \Lambda_b$ for the biasing configurations II and III are, respectively $0.11$ and $0.21$. With $\Lambda_b \approx 0.26$, we end up with $\alpha_0=0.42$ and $0.81$, respectively, for the biasing configurations II and III, reflecting the fact that, as expected, the configuration II belongs to the asymmetric range, $\alpha<0.5$, while the configuration III belongs to the symmetric one, $\alpha>0.5$. Furthermore, zero-field current singularities were observed in the junction current-voltage curves that remind of the resonant fluxon motion; however, due to the phase torque induced by the fluxoid quantization that mimic the effect of an external magnetic field, we believe that they are better ascribed to the asymmetric fluxon propagation of Fiske-type resonances, observed in LJTJ\cite{fiske}.

\noindent We can now comment on the transverse MDPs in Fig.~\ref{exp}(b) in which drastic $I_c$ changes are found but only for negative $H_Z$ values. In this case the externally induced circulating currents become increasingly important and, as discussed in the Section II, we have to consider the effects of the energy minimization with $\alpha$ being a degree of freedom bonded to the $[0,1]$ interval. It can be shown that, for generic $h_{r,l}$ values, the junction energy in Eq.(\ref{energy}) is minimum when $2 \alpha \Lambda=1+2h_{rad}/\iota$ (we note that $h_{app}=H_Y=0$ in this situation). For concord $h_{rad}$ and $\iota$, the parameter $\alpha$ is squeezed to its upper value $1$, while for $h_{rad}<-\iota/2$ it is bounded to zero; however, with the ratio $2h_{rad}/\iota \in [-1,0]$, the free parameter $\alpha$ will assume the value(s) in the range $[0,1]$ that minimize the energy. It is possible that, for a given $h_{rad}$ ($H_Z$ in the experiments), two values of $\iota$ ($I$) exist that correspond to different junction energy levels. The coexisting states are evident in the transverse MDPs for for $H_z$ in the range $[-0.5,-0.25]H_{Z,c}$. Therefore, we believe that the sudden changes in $I_c(H_Z)$, that are absent in the in-plane MDPs, correspond to abrupt modifications of the phase profile $\phi(x)$ along the LJTJ and indicate that the energy minima correspond to quite different $\alpha$ values. Numerical simulations are planned to understand the experimental MDPs at a quantitative level. Finally, the appearance of displaced linear slopes for transverse magnetic fields larger than the critical value constitutes one more indication that the phase twist is increasing with the circulating currents, as suggested by Eq.(\ref{constraint}). 

\subsection{Signal-to-noise ratio}

\begin{figure}[tb]
\centering
\includegraphics[width=7.5cm]{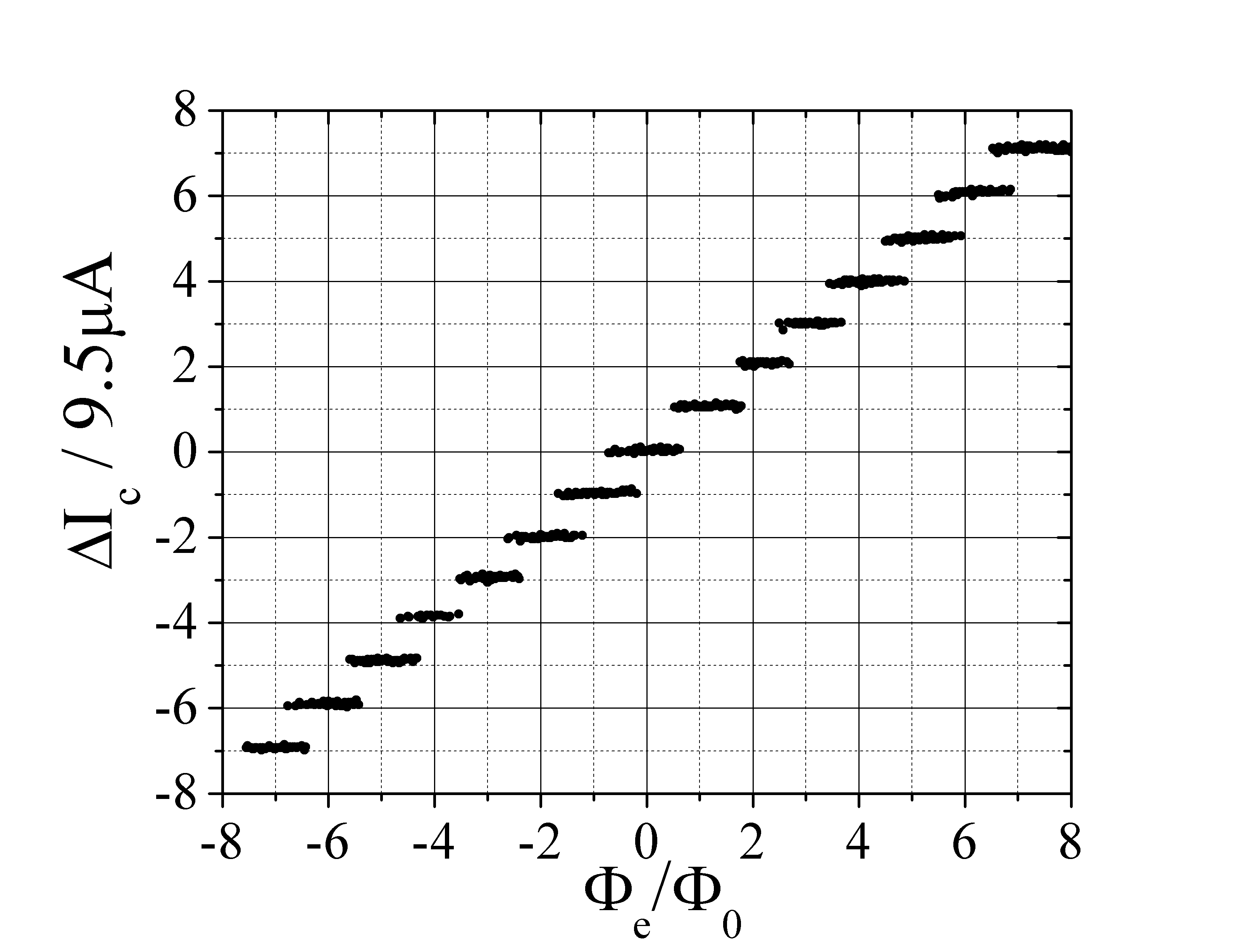}
        \caption{Variation of the critical current as a function of the magnetic flux with which the loop was field-cooled through the superconducting transition temperature. $H_Y=H_Z=0$ and $T=4.2\,$K.\label{calibration}}
\end{figure}

With our devices a very large signal-to-noise ratio can be achieved in the detection of magnetic flux quanta trapped in the loop. Fig.~\ref{calibration} shows how the zero-field critical current changes of anyone of the two LJTJs on top of the rectangular loop when the system is cooled through the NS transition in the presence of a transverse magnetic field which is incremented by steps corresponding to a small fraction of the magnetic flux quantum, $\Delta \Phi_e=0.02\,\Phi_0$. Once the transverse field is removed, the quantized levels of the critical currents with $\Delta I_c\approx 9.5\, \mu$A are clearly visible out of the $\pm 1 \, \mu$A rms current noise from the thermal fluctuation of the critical current.

\noindent As the MDPs of Fig.~\ref{exp} indicate, as far as $|h_e| \leq |h_{max}|$, the slope of the positive and negative critical currents, respectively, $I_{c}^+$ and $I_{c}^-$, is the same. Consequently, any circulating current $I_{cir}$ modulates $I_{c}^+$ and $I_{c}^-$ concordly, meaning that the offset current $I_c^{off} \equiv I_{c}^+ + I_{c}^-$ changes twice faster. The root mean square noise of $I_c^{off}$ is $\sqrt{2}$ times larger than that of a single critical current, meaning that the signal-to-noise ratio is enhanced by a factor $\sqrt{2}$. In addition to this, since the change with temperature of $I_{c}^+$ is numerically the same but opposite to that of $I_{c}^-$, one more advantage of the current offset, as compared to just one critical current, is its much reduced sensitivity to any temperature drift that might occur during the measurements. The data shown in Fig.~\ref{calibration} refer to $I_c^{off}/2$.

\noindent An accurate value of the loop inductance can be obtained from Eq.(\ref{Delta}):
$$L_{loop} =  \frac{g_i \Phi_0}{\Delta I_{c}}.$$

\noindent For the rectangular loop with the bias configuration I ($g_i=0.33$) we find $L_{loop}=72\, $pH. For the circular loop with the same biasing configuration ($g_i=0.35$), we measured $\Delta I_{c}=5.2\, \mu$A corresponding to $L_{loop}=140\,$pH. 

\noindent The superposition of the levels is due to the non-adiabaticity of the thermal transitions\cite{zurek2} and the transition from one level to the next follows a Gaussian probability law\cite{PRB12}. In fact, during the normal-to-superconducting transition the loop temperature changed at a rate of $5 \times 10^{3}\,$K/s. Indeed, this method is strongly inspired by the results found in our investigation of the spontaneous fluxoid formation in superconducting loops based on the detection of the persistent currents circulating around a hole in a superconducting film, when one or more fluxoids are trapped inside the hole\cite{PRB09}. 

\section{CONCLUSIONS}

\noindent In this paper we have revisited the theory of the self-field effects that characterize the long Josephson tunnel junctions and made them interesting for the investigation of non-linear phenomena. Our analysis goes beyond the previous works in  two ways: (i) it takes into account the different inductances per unit length of the electrodes forming the junction and (ii) it provides the boundary conditions for the most general junction biasing configuration. We applied the theory to the specific case of long Josephson tunnel junctions with not simply connected electrodes. Apart from their intriguing physical properties, the interest for LJTJs built on a superconducting loop stems from the fact that they were successfully used to detect trapped fluxoids in a {\it cosmological} experiment aimed to study the spontaneous defect production during the fast quenching of a superconducting loop through its normal-to superconducting transition temperature\cite{PRB09}. We found that the single-valuedness of the phase $\phi_1$ of the order parameter of the bottom (or top) superconducting electrode (fluxoid quantization) gives raise to a variety of unexpected non-linear phenomena when coupled to the sine-Gordon equation for the phase difference $\phi_2-\phi_1$ of the order parameters in the superconductors on each side of the tunnel barrier. The principle of energy minimization was also invoked to determine the possible states of the system. We have focused on static phenomena such as the reduction of the junction critical current and its dependence on magnetic fields applied in and out of the loop plane. Nevertheless, also the dynamic properties, such as the propagation of non-linear waves, are expected to be drastically affected by the doubly connected electrode(s). Our experiments unambiguously corroborate the analytical findings and provide hints to implement the modeling. Future work should go in the direction of investigating the consequences of the fluxoid quantization for small and intermediate length Josephson tunnel junctions and on the dynamic properties of long junctions. 

\section*{ACKNOWLEDGMENT}
\noindent VPK acknowledges the financial support from the Russian Foundation
for Basic Research under the grants 11-02-12195 and 11-02-12213.

\end{document}